\documentclass[sigconf]{acmart}

\AtBeginDocument{%
  }

\copyrightyear{2025}
\acmYear{2025}
\setcopyright{cc}
\setcctype{by}
\acmConference[CHI '25]{CHI Conference on Human Factors in Computing Systems}{April 26-May 1, 2025}{Yokohama, Japan}
\acmBooktitle{CHI Conference on Human Factors in Computing Systems (CHI '25), April 26-May 1, 2025, Yokohama, Japan}\acmDOI{10.1145/3706598.3714229}
\acmISBN{979-8-4007-1394-1/25/04}

\acmConference[CHI '25]{ACM CHI Conference on Human Factors in Computing Systems}{April 26--May 1,
  2025}{Yokohama, JP}

\usepackage{xspace}
\xspaceaddexceptions{\}}
\usepackage{enumitem}
\usepackage{subcaption}
\usepackage{amsmath}
\usepackage{soul}
 \usepackage{acmart-taps}

\graphicspath{{Figures/}} % where to search for the images

\newcommand{\legendbox}[1]{%
\textcolor{#1}{\rule{\fontcharht\font`X}{\fontcharht\font`X}}%
}
\definecolor{youngeradults}{HTML}{F28E2B}
\newcommand{\yacbox}{\legendbox{youngeradults}\xspace}
\newcommand{\placbox}{\legendbox{black}\xspace}

\newcommand{\taskAn}{Find~\textsc{Anomalies}\xspace}
\newcommand{\taskCl}{Find~\textsc{Clusters}\xspace}
\newcommand{\taskCo}{Find~\textsc{Correlation}\xspace}
\newcommand{\taskDe}{Compute~\textsc{Derived}~Value\xspace}
\newcommand{\taskDi}{Characterize~\textsc{Distribution}\xspace}
\newcommand{\taskEx}{Find~\textsc{Extremum}\xspace}
\newcommand{\taskFi}{\textsc{Filter}\xspace}
\newcommand{\taskOr}{\textsc{Order}\xspace}
\newcommand{\taskRa}{Determine~\textsc{Range}\xspace}
\newcommand{\taskRe}{\textsc{Retrieve}~Value\xspace}

\newcommand{\metAcc}{accuracy\xspace}
\newcommand{\metTime}{time\xspace}

\newcommand{\visBar}{\textit{Bar}\xspace}
\newcommand{\visLine}{\textit{Line}\xspace}
\newcommand{\visPie}{\textit{Pie}\xspace}
\newcommand{\visScatter}{\textit{Scatterplot}\xspace}
\newcommand{\visTable}{\textit{Table}\xspace}

\definecolor{visBar}{HTML}{648FFF}
\newcommand{\barcbox}{\legendbox{visBar}\xspace}
\definecolor{visLine}{HTML}{785EF0}
\newcommand{\linecbox}{\legendbox{visLine}\xspace}
\definecolor{visPie}{HTML}{DC267F}
\newcommand{\piecbox}{\legendbox{visPie}\xspace}
\definecolor{visScatter}{HTML}{FE6100}
\newcommand{\scattercbox}{\legendbox{visScatter}\xspace}
\definecolor{visTable}{HTML}{FFB000}
\newcommand{\tablecbox}{\legendbox{visTable}\xspace}

\newcommand{\iDotE}{i.\,e.,\xspace}
\newcommand{\eDotG}{e.\,g.,\xspace}

\begin{document}

\title[Toward Filling a Critical Knowledge Gap]{Toward Filling a Critical Knowledge Gap: Charting the Interactions of Age with Task and Visualization}

\author{Zack While}
\email{zwhile@cs.umass.edu}
\orcid{0000-0002-9114-3984}
\affiliation{%
  \institution{University of Massachusetts Amherst}
  \city{Amherst}
  \state{Massachusetts}
  \country{USA}
}
\author{Ali Sarvghad}
\email{asarv@cs.umass.edu}
\orcid{0000-0003-3718-7043}
\affiliation{%
  \institution{University of Massachusetts Amherst}
  \city{Amherst}
  \state{Massachusetts}
  \country{USA}
}

\begin{abstract}
    We present the results of a study comparing the performance of younger adults~(YA) and people in late adulthood~(PLA) across ten low-level analysis tasks and five basic visualizations, employing Bayesian regression to aggregate and model participant performance. We analyzed performance at the task level and across combinations of tasks and visualizations, reporting measures of performance at aggregate and individual levels. These analyses showed that PLA on average required more time to complete tasks while demonstrating comparable accuracy. Furthermore, at the individual level, PLA exhibited greater heterogeneity in task performance as well as differences in best-performing visualization types for some tasks. We contribute empirical knowledge on how age interacts with analysis task and visualization type and use these results to offer actionable insights and design recommendations for aging-inclusive visualization design. We invite the visualization research community to further investigate aging-aware data visualization. Supplementary materials can be found at~\url{https://osf.io/a7xtz/}.
\end{abstract}




\maketitle

\section{Introduction}
Does our empirically-driven knowledge of visualization design, use, and evaluation apply to all demographics?  This work is guided by and investigates this question by focusing on people in the late adulthood (PLA) stage of development~\cite{berk2022development}. With age often comes a gradually progressing decline in our perception, cognition, and motor skills ~\cite{roberts2016perception, kiss2001age,johnson2017designing}. Hence, it is plausible that aging may affect our ability to use data visualizations. Nevertheless, our existing empirical knowledge of PLA's use of visualizations, issues and challenges, and strategies to alleviate them is considerably scarce~\cite{backonja2016visualization,le-2016-eval,while2024gerontovis}. This knowledge gap becomes even more critical in light of the fast-growing population of PLA and the rapid emergence of data-driven tools and technologies that use visualization to
display various tracked aspects of PLA's health, physical activity, and daily routines (\eDotG~\cite{fanning2018mobile,gualtieri2016can,kimura2019modifiable, tapia2016designing}) as well as to make data-driven decisions (\eDotG~\cite{garcia2012using, galesic2009using, ruiz2013communicating,price2016effects}). 
According to the US Census Bureau, by the year 2035, the PLA population in the US (77 million, 21.4\%) will start exceeding the population of children (75.5 million, 21\%)~\cite{ortman2014aging}. 
With these observations in mind,  we argue that it is imperative to reassess and extend our knowledge of visualization design and use for PLA. Similarly, Shneiderman et al.~\cite{shneiderman2016grand} argue that designing information interfaces for PLA is one of the current grand challenges in the human-computer interaction (HCI) research field. In this work, we report and discuss the results of a user study and Bayesian modeling analysis aimed at extending our empirical knowledge of PLA's visual analysis performance and design considerations. 

The central objective of this research is to investigate and demonstrate how aging might affect the ability to perform visual analysis tasks, with existing guidelines derived from studies involving younger participants~\cite{while2024gerontovis}. We hope to alert the visualization research community to the potential impact of age on visualization effectiveness and to provide empirical evidence demonstrating how existing visualization knowledge may not universally apply to all age groups. This study emphasizes the need for a more nuanced understanding of how different demographics interact with visualizations, challenging the assumption that principles validated with younger populations are automatically suitable for PLA.
To this end, we aimed to compare, contrast, and learn about the possible differences between YA and PLA's performance. 
We replicated Saket et al.'s crowdsourced study~\cite{saket2018task} in which they examined the relationship between ten primary visual data analysis tasks (\eDotG identifying outliers) and five basic data presentation methods (\eDotG a Scatterplot) according to the performance (\metAcc, \metTime) of 203 participants aged between 25 to 40. One of the current challenges in broader aging research is robustly defining who would be considered PLA, so we followed prior work (\eDotG~\cite{le2014design}) by recruiting people age 60 and older to have a sample with reasonable variance. The two studies are identical except for the participants' ages and the number of pre-study survey questions (see \autoref{sec:study-1} for details). 

Recent work in HCI has advocated for the use of Bayesian statistics to help the field accumulate knowledge (via improving priors over time with increasing amounts of data) and arrive at more robust results with smaller sample sizes~\cite{kay2016researcher}. While able to model patterns at the population level, these approaches are especially well-suited to model individual differences, which focus on how both user-specific factors may impact design as well as how performance can vary between participants in a study~\cite{davis2022risks}. As of late, this approach to design has gained increasing traction as a design consideration in both HCI and visualization~\cite{liu2020survey,davis2022risks,ge2023calvi}.  

Thus, we used Bayesian linear modeling for the \metAcc and \metTime data to model performance differences at the individual and the population levels.  Next, using these models' posterior distributions, we generated performance data for 12000 \textit{counterfactual}~\cite{mcelreath2018statistical} (simulated) participants to increase the likelihood of better capturing the wide variance that may exist in reality. This strengthened the robustness and generalizability of our analyses by exploring patterns that may otherwise be hidden in the empirical data~\cite{mcelreath2018statistical}.
Sampling new participants and analyzing the \metTime data, we found that PLA consistently required more time than YA across all combinations of tasks and visualizations (Avg. 56\% difference). At the task level, PLA were nearly as accurate as YA (Avg. 6\% difference). 
Furthermore, the differences between the two groups' \metAcc were rather sporadic across the combinations of tasks and visualizations. 
The most notable difference in PLA's \metAcc occurred for \taskDi, while the greatest differences in \metTime were for \taskAn, \taskCo, and \taskFi. They also consistently had Line charts as the lowest-ranked visualization for \metAcc across several tasks. Additionally, PLA's \metAcc using a Table was noticeably better than YA's for many tasks.

Our work contributes to advancing our empirical understanding of the complex interplay between age, visualization design choices, and visual analysis tasks. 
We discuss the performance differences between and within age groups and posit factors that may possibly contribute to inter- and intra-group variations. Finally, grounded in the outcomes of our analysis, we discuss practical considerations and implications for designing visualizations with PLA in mind.

\section{Related Work}
In this section, we review prior work investigating visual data analysis focusing on people in late adulthood (PLA). We split this into two distinct areas. First, \textit{psychophysical-focused work}, which is more broadly about how physiological factors can affect visual analysis performance. Second, \textit{application-focused work}, which focuses on effectively applying visualization to a specific domain for PLA.

\subsection{Psychophysical-Focused Work}
Information visualization is an established scientific field with extensive existing literature. However, there is a notable gap in the number of studies empirically investigating perceptual and cognitive aspects of visualization for PLA compared to younger adults (YA); recent work has also pointed to this gap in the literature~\cite{backonja2016visualization, while2024gerontovis}. 

Closest to our study, Le et al.~\cite{le2014elementary} investigated the differences in comparison and proportion judgment tasks between PLA (60 years or older) and YA (below 60 years old). Recruiting a total of 202 participants in two remote studies, they measured participants' accuracy and performance time using Bar, Stacked Bar, and Pie charts. They found that, on average, PLA were 4.09 and 3.66 seconds slower across comparison and proportion judgment tasks. PLA were also less accurate; however, the differences were not statistically significant. 
While et al.~\cite{while2024glanceable} investigated how aging affects the amount of time needed to compare data on a glanceable smartwatch visualization, finding strong evidence that PLA required more time than YA across multiple amounts of depicted data.
In a recent study, van Weert et al.~\cite{van2021preference}  
compared the understanding verbatim and gist knowledge of YA ($<65$, $n=219$) and PLA ($\geq 65$, $n=227$) when visually presented with health-risk information. According to Fuzzy Trace Theory~\cite{reyna1995fuzzy}, verbatim knowledge pertains to the ability to capture the exact words, numbers, or images, while gist knowledge pertains to higher-level interpretation of a stimulus. They found that, at the same level of graph literacy and numeracy,  PLA's verbatim knowledge of visualizations was less accurate than those of their younger counterparts; furthermore, they suggested that Bar graphs and Tables would positively impact PLA's verbatim and gist understanding of graphs.
Similarly, Hawley et al.~\cite{hawley2008impact} found that adults age 60 and older were significantly less likely to showcase sufficient verbatim knowledge ($n=344$) and gist knowledge ($n=485$) compared to adults less than 40 years old and adults 41-59 years old.
All of these studies are either adjacent to visualization or focus on very specific types of visualizations, while our study covers five visualizations each across ten tasks.

While et al.~\cite{while2024gerontovis} established the research area of GerontoVis, arguing for the importance of work at the intersection of visualization and aging as well as providing an outline of existing work and future challenges. Shao et al.~\cite{shao2024does} performed a literature review, aggregating knowledge from both the visualization and psychophysics literature relevant to how aging can affect visualization performance. They then created a linear model which, assuming the use of a 14-inch HD laptop display, found that age-inclusive visualizations can be created given a high enough amount of contrast and low enough spatial frequency of visual artifacts. They argue that modern digital visualizations should be able to fit this criteria, however applying this approach to other screen sizes would require further investigation. 
Price et al.~\cite{price2016effects} investigated the use of visualizations for enabling PLA to choose among Medicare plans, finding that visualizations reduced their working memory demand, consequently improving the accuracy of their decisions. 
A considerable amount of prior research has also investigated the topic of universal accessibility and usability of visualizations. The main goal of these works was to understand how people with disabilities can access and use visualizations (\eDotG~\cite{lee2018data,plaisant2005information, vartak2015seedb,007_zou2016chartmaster}). However, recent work has cautioned against equating aging work in HCI with accessibility~\cite{knowles2021harm}, while others have discussed how accessibility work is not well-suited on its own to cover the progressive, compounding ways that changes with aging can impact visualization performance~\cite{while2024gerontovis}.

Our review of the psychophysical-focused work at the intersection of visualization and aging showed a lack of \textit{empirical examination and evaluation} of existing visualization design knowledge with respect to PLA. 
Hence, this work aims to narrow the current knowledge gap through an empirical study and detailed analysis.

\subsection{Application-Focused Work}
In addition to research focusing on psychophysical performance in the context of aging and visualization, there is also a body of application-focused work that examines the use of visualizations as a means of supporting PLA in executing specific real-world tasks, \eDotG tracking and understanding attributes like health~\cite{cajamarca2023understanding} and physical activity~\cite{vargemidis2023performance}. 
We focus on two main types of papers in these areas: those focusing on visualization design for PLA as well as those focusing on PLA's experiences using visualizations.

Whitlock \& McLaughlin~\cite{whitlock} ran a usability analysis for three different blood glucose tracking mobile apps for PLA, noting that the visualizations had axis labels that were too small, Line graphs with too-narrow edges, and poor choices of color and contrast. Morey et al.~\cite{morey-heart-failure} used a heuristic analysis to analyze two heart health apps, also observing low color contrast, small fonts, and a cluttered presentation, with many users noting difficulty in understanding the data being presented while acknowledging the advantages of having the information available. Le et al.~\cite{Le2012-integrated} designed charts for PLA's wellness visualizations using cognition and graphical comprehension theory, while Reeder et al.~\cite{Reeder2014-gk} used participatory design to collaboratively create wellness visualizations with PLA for them and their stakeholders. In order to help practitioners and researchers working in the area of health visualization for PLA, Le et al.~\cite{le-2016-eval} provide a comparison of three different evaluation methods, concluding that each type provides perspectives in one major evaluation criteria. A study by Ribeiro et al.~\cite{ribeiro2017souschef} showed visualization could assist PLA in the selection of meal plans based on their tracked health and activity data; further evidence has been shown in its ability to aid in fall prevention~\cite{hamm2017fall} and diabetes self-management~\cite{burford2019small}. 
Huh et al.~\cite{Huh2013-rp} found that PLA's low confidence in understanding presented wellness information obtained via computers created difficulties in their perceived ability to properly interpret and act upon it. Observing a similar technology literacy issue, Le et al.~\cite{le-2018-understanding} recommended providing different access methods such as paper printouts. Participants in those semi-structured interviews also expressed interest in having context provided for various charts in order to better interpret the results for their health needs, which was further echoed in other studies~\cite{pham2012effects}. In a participatory design study by Ahmed et al.~\cite{ahmed2019visualization}, PLA preferred using a Bar chart instead of a Line chart for temporal data and enjoyed using a familiar stoplight metaphor for color-coding heart pace alerts.

Because of these works' focus on visualizations depicting specific types of data (\eDotG health and wellness), some results may not generalize well to a more broad understanding of aging's impact on visualization performance.  As a result, approaching age-inclusive visualization from a design-centric perspective is still understudied, further showcasing the necessity for research in this direction.

\section{Study Methodology and Design}
\label{sec:study-1}

This work aimed to investigate how aging might affect visualization performance. To this end, we replicated a study by Saket et al.~\cite{saket2018task} that provided performance-based comparisons (\metAcc, \metTime) of five basic visualizations (\visBar, \visLine, \visPie, \visScatter, \visTable) across ten fundamental visual data analysis tasks (detailed in \autoref{taks}). We chose this particular study for two main reasons. 
First, it provided data on visualization performance derived from a younger cohort (ages 25-40).
Secondly, the comprehensive availability of details regarding the study’s design, procedures, execution, and the collected data made it feasible to more accurately recreate the study with PLA. 
From this point forward, we will refer to it as the \textit{prior study}.

The design and execution of our study and the prior study were identical except for two key differences: (1) the participants' age, which was 25-40 in the prior study and 60+ in our study, and (2) an increased number of pre-study questionnaire questions from three (age, gender, level of familiarity with visualizations) to four, additionally collecting participants' highest level of education. Our methodology closely aligns with \textit{conceptual replication}~\cite{brandt2014replication, crandall2016scientific, stroebe2014alleged}, a replication study type in which the same research questions are examined as in the prior study but with \textit{controlled changes} in methods, participants, or settings.  Conceptual replication differs from direct replication by aiming to test and expand the generalizability of previous findings rather than seeking to confirm the original results. While et al.~\cite{while2024glanceable} recently used a similar methodology to replicate a study by Blascheck et al.~\cite{blascheck2018glanceable} on glanceable visualizations for PLA.

The following subsections provide a detailed description of the study design, procedure, online platform, and execution.

\subsection{Participants}
\label{subsec:study-1-participants}
 We used the Prolific~\cite{Prolific80:online} crowdsourcing platform to recruit 200 US-based PLA (compared to 203 YA in the prior study) and run the study. Participants' selection criteria were age 60+, an approval rate of 95\% or higher, and the use of a desktop computer. We did not opt for mobile or tablet users in order to minimize the effects of screen size on participants' performance. 
 Each participant could only take the study once, and sessions took anywhere from 30 to 45 minutes to complete. Each participant received \$9 as compensation. 
 \autoref{tbl:demographics} shows the detailed demographic information of participants collected in the pre-study questionnaire; note that the visualization familiarity was given on a rating 1-10, where 1 indicated \textit{no familiarity} and 10 indicated \textit{high familiarity}. 
We observed no noticeable impact on participants' accuracy and time due to any of the demographic categories, with more details provided in the supplementary materials.
All participants in our online study had normal or corrected-to-normal vision, with further details provided in \autoref{tbl:vision}.
 Notably, while 200 participants were recruited for this study, 52 participants had results removed due to poor and/or incomplete performance (detailed in \autoref{sec:data-prep}), resulting in a final participant pool of 148 PLA compared to 180 YA in the prior study. 

 \begin{figure*}[ht]
    \centering
    \captionsetup{type=table}
    \captionof{table}{Demographic information for the 148 PLA participants in the study: (a) age range, (b) gender, (c) visualization familiarity (scale 1-10), and (d) education level. Reported values are the percentage of participants of a given demographic attribute.}
    \label{tbl:demographics}
    \begin{subfigure}{0.2\textwidth}
        \centering
        \caption{Age Range}
        \resizebox{\textwidth}{!}{%
        \begin{tabular}{rr}
        \hline
        \textbf{Age Range} & \textbf{\% of Participants} \\ \hline
        60-65 & 63\% \\
        66-70 & 24\% \\
        71-75 & 9\% \\
        76+ & 4\% \\ \hline
        \end{tabular}%
        }
    \end{subfigure}
    \hfill
    \begin{subfigure}{0.23\textwidth}
        \centering
        \caption{Gender}
        \resizebox{\textwidth}{!}{%
        \begin{tabular}{rr}
        \hline
        \textbf{Gender} & \textbf{\% of Participants} \\ \hline
        Female & 50\% \\
        Male & 50\% \\
        Prefer Not to Say & 0\% \\
        Other & 0\% \\ \hline
        \end{tabular}%
        }
    \end{subfigure}
    \hfill
    \begin{subfigure}{0.235\textwidth}
        \centering
        \caption{Visualization Familiarity}
        \resizebox{\textwidth}{!}{%
        \begin{tabular}{rr}
        \hline
        \textbf{Vis. Familiarity} & \textbf{\% of Participants} \\ \hline
        Low (1-3) & 10\% \\
        Medium (4-7) & 38\% \\
        High (8-10) & 52\% \\ \hline
        \end{tabular}%
        }
    \end{subfigure}
    \hfill
    \begin{subfigure}{0.23\textwidth}
        \centering
        \caption{Education Level}
        \resizebox{\textwidth}{!}{%
        \begin{tabular}{rr}
        \hline
        \textbf{Education Level} & \textbf{\% of Participants} \\ \hline
        High School & 22\% \\
        Undergraduate & 36\% \\
        Graduate & 37\% \\
        Other & 4\% \\ \hline
        \end{tabular}%
        }
    \end{subfigure}
    \vspace{1em}
\end{figure*}

 \begin{figure*}[ht]
    \centering
    \captionsetup{type=table}
    \captionof{table}{Vision-related information for the 148 PLA participants in the study: (a) use of corrective glasses, (b) use of reading glasses, and (c) reported corrected-to-normal vision impairments. Reported values are the percentage of participants of a given attribute.}
    \label{tbl:vision}
    \begin{subfigure}{0.285\textwidth}
        \centering
        \caption{Corrective Glasses}
        \resizebox{\textwidth}{!}{%
        \begin{tabular}{rr}
        \hline
        \textbf{Corrected Impairment} & \textbf{\% of Participants} \\ \hline
        Farsighted & 31\% \\
        Nearsighted & 43\% \\
        \textit{Not Applicable} & 26\% \\ \hline
        \end{tabular}%
        }
    \end{subfigure}
    \hspace{1.5em}
    \begin{subfigure}{0.225\textwidth}
        \centering
        \caption{Reading Glasses}
        \resizebox{\textwidth}{!}{%
        \begin{tabular}{rr}
        \hline
        \textbf{Reported Use} & \textbf{\% of Participants} \\ \hline
        Yes & 69\% \\
        No & 31\% \\ \hline
        \end{tabular}%
        }
    \end{subfigure}
    \hspace{1.5em}
    \begin{subfigure}{0.285\textwidth}
        \centering
        \caption{Vision Impairments}
        \resizebox{\textwidth}{!}{%
        \begin{tabular}{rr}
        \hline
        \textbf{Corrected Impairment} & \textbf{\% of Participants} \\ \hline
        Cataracts & 14\% \\
        Glaucoma & 0\% \\
        Macular Degeneration & 0\% \\
        \textit{Not Applicable} & 86\% \\ \hline
        \end{tabular}%
        }
    \end{subfigure}
\end{figure*}

\subsection{Study Platform and Dataset}
\label{subsec:study-1-study-platform}
We used the source code and the dataset used in the prior study, which the authors provided us with, to recreate the experiment precisely. Hence, the flow (pre-study questionnaire $\rightarrow$ training and warm-up tasks $\rightarrow$ main experiment $\rightarrow$ follow-up questions), randomization of task assignments (simple randomization technique), and counterbalancing were identical between the two studies. We also used the same two tabular datasets, \textit{Cars} and \textit{Movies}, which were used in the prior study. The Cars dataset has information for 407 cars (\eDotG price, model, miles per gallon), and the Movies dataset includes details for 335 movies (\eDotG genre, length) released from 2007 to 2012.
Each participant was randomly-assigned one task, for which they would answer a total of 30 questions (5 visualizations each used for 6 different questions), in addition to the two aforementioned ``test'' questions. The prior study had additional questions regarding visualization preference for a given task, however this type of data was beyond the scope of our focus.

\subsection{Analysis Tasks}\label{taks}
Data analysis tasks performed by our participants were identical to those performed by those in the prior study, which were based on the set of ten low-level analysis tasks described by Amar et al.~\cite{amar2005low}.  It is of note that questions for a given task were the same, regardless of the visualization that was used to present the data; some types of tasks may better lend themselves to certain visualizations (and vice versa), which was one of the major research questions and results of the prior study~\cite{saket2018task} and is also a component of our study. The rest of this section consists of definitions and examples of each task:\\

\noindent\textbf{Find Anomalies.} 
For this task, participants were asked to identify any irregular (\eDotG a car with negative miles per gallon) data points within a depicted dataset. For example, \textit{which car manufacturer appears to build very high-consumption engines?}  It is of note that the prior study created these anomalies manually~\cite{saket2018task}; however, because we used the exact data and software as used in the prior study, our anomalies are the exact same and thus allow for a fair comparison between groups.\\

\noindent\textbf{Find Clusters.} For this task, participants were asked to identify and count the number of groups based on a shared attribute between data points. For example, \textit{how many different car manufacturers are shown in the chart below?}\\

\noindent\textbf{Find Correlation.} For this task, participants were asked whether two presented data attributes were correlated. For example, \textit{is there a strong correlation between the average number of cylinders and miles per gallon consumption?}\\

\noindent\textbf{Compute Derived Value.} 
For this task, participants were asked to calculate a summary measure for a set of data values. For example, \textit{what is the average price of eight-cylinder cars?}\\

\noindent\textbf{Characterize Distribution.} For this task, participants were asked to describe how the data values were distributed. For example, \textit{what percentage of the cars have a price greater than 80k?}\\

\noindent\textbf{Find Extremum.} For this task, participants were asked to identify data points with noticeable outlier values for a given attribute. For example, \textit{what is the car with the highest price tag?}\\

\noindent\textbf{Filter.} For this task, participants were asked to find data points that met a predefined set of criteria. For example, \textit{which car types have price ranging from 30k to  40k?}\\

\noindent\textbf{Order.} For this task, participants were asked to rank data values based on one of their ordinal attributes. For example, \textit{which of the following options shows the correct sequence of cars, ordered from highest to lowest prices?}\\

\noindent\textbf{Determine Range.} For this task, participants were asked to compute the numerical extent of a set of data points. For example, \textit{what is the range of the number of cylinders for cars shown in this graph?}\\

\noindent\textbf{Retrieve Value.} For this task, participants were asked to determine the value of a specified data point. For example, \textit{what is the horsepower of Buick Regal 1996 model?}

\subsection{Visualizations}
\label{subsec:study-1-vis-and-user-tasks}

\begin{figure*}[ht]
\captionsetup[subfigure]{justification=centering}
\centering
\subfloat[\taskRe, \visScatter][\taskRe,\\\visScatter]{
\includegraphics[width=0.188\textwidth]{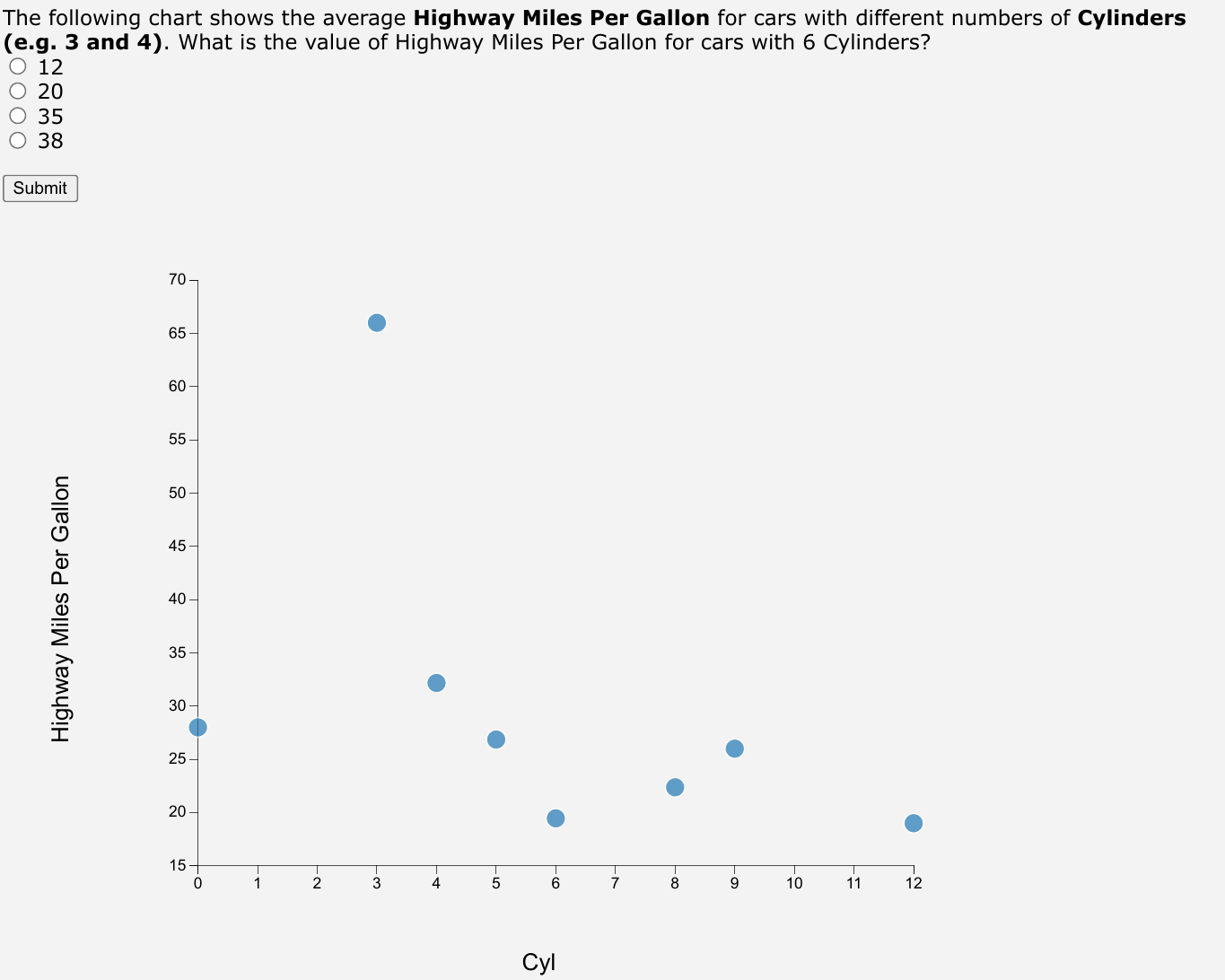}
\Description{Example questions with a scatterplot.}
}
\subfloat[\taskAn,\\\visPie]{
\includegraphics[width=0.205\textwidth]{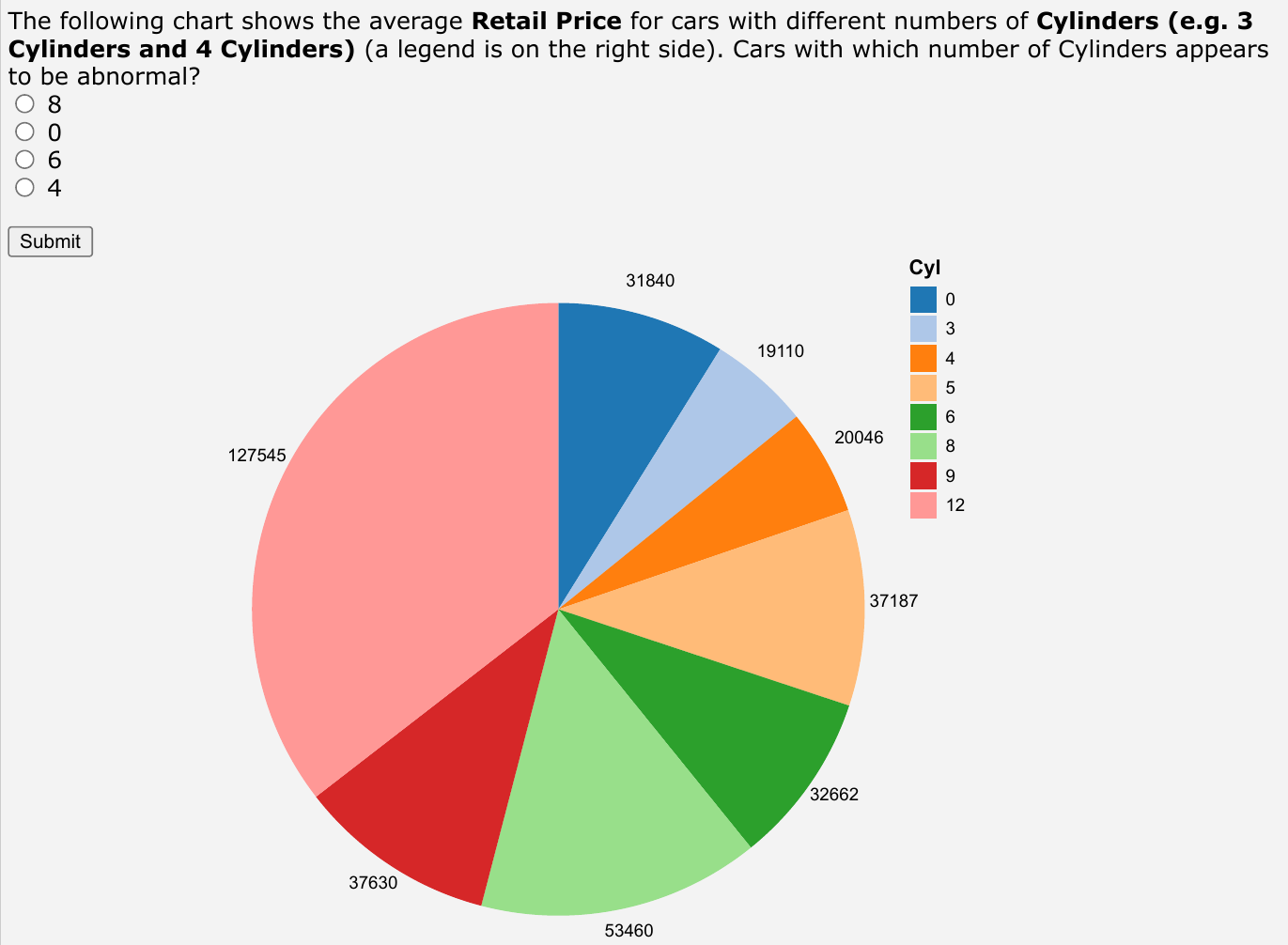}
\Description{Example questions with a pie chart.}
}
\subfloat[\taskFi,\\\visBar]{
\includegraphics[width=0.189\textwidth]{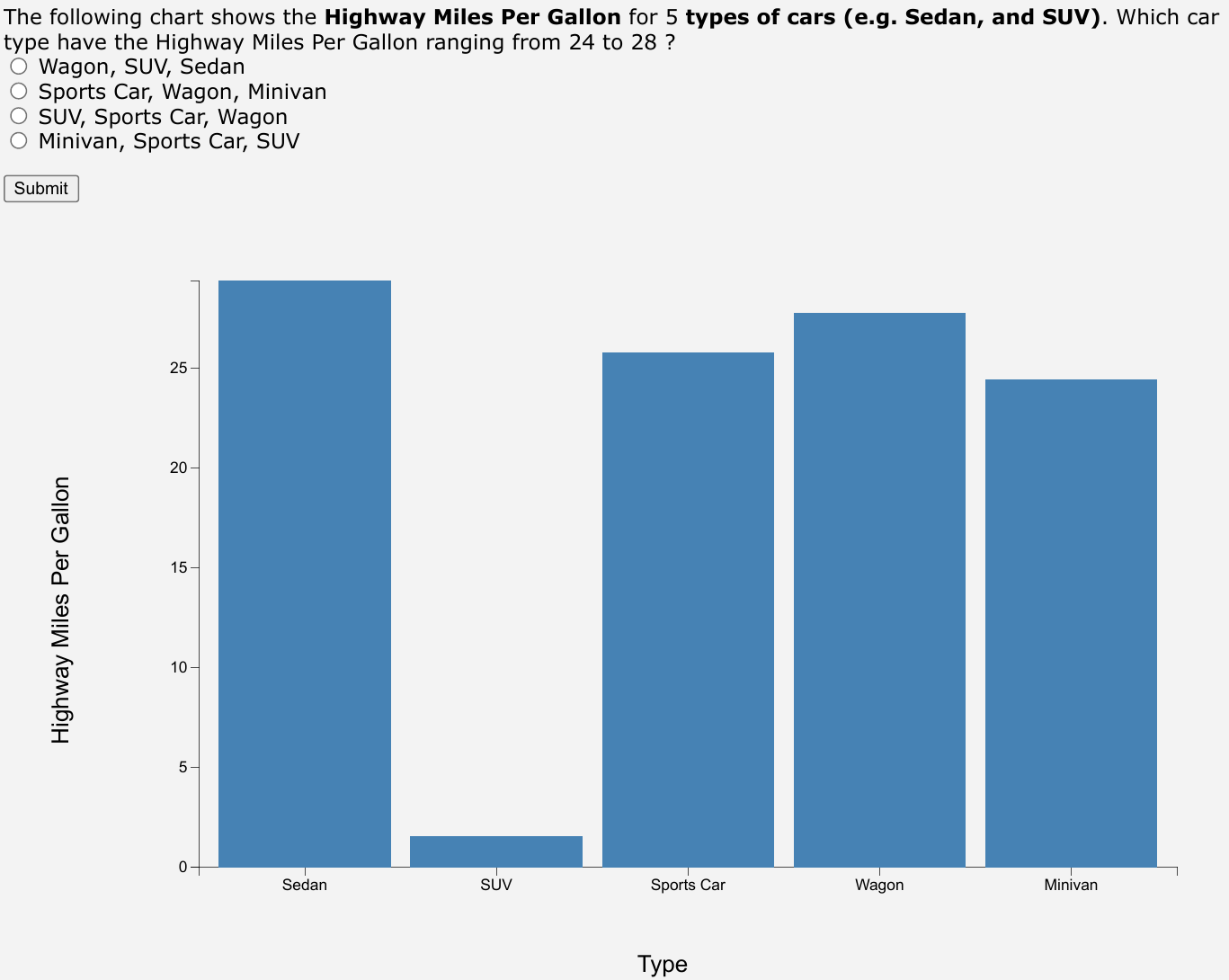}
\Description{Example questions with a bar.}
}
\subfloat[\taskRa,\\\visLine]{
\includegraphics[width=0.188\textwidth]{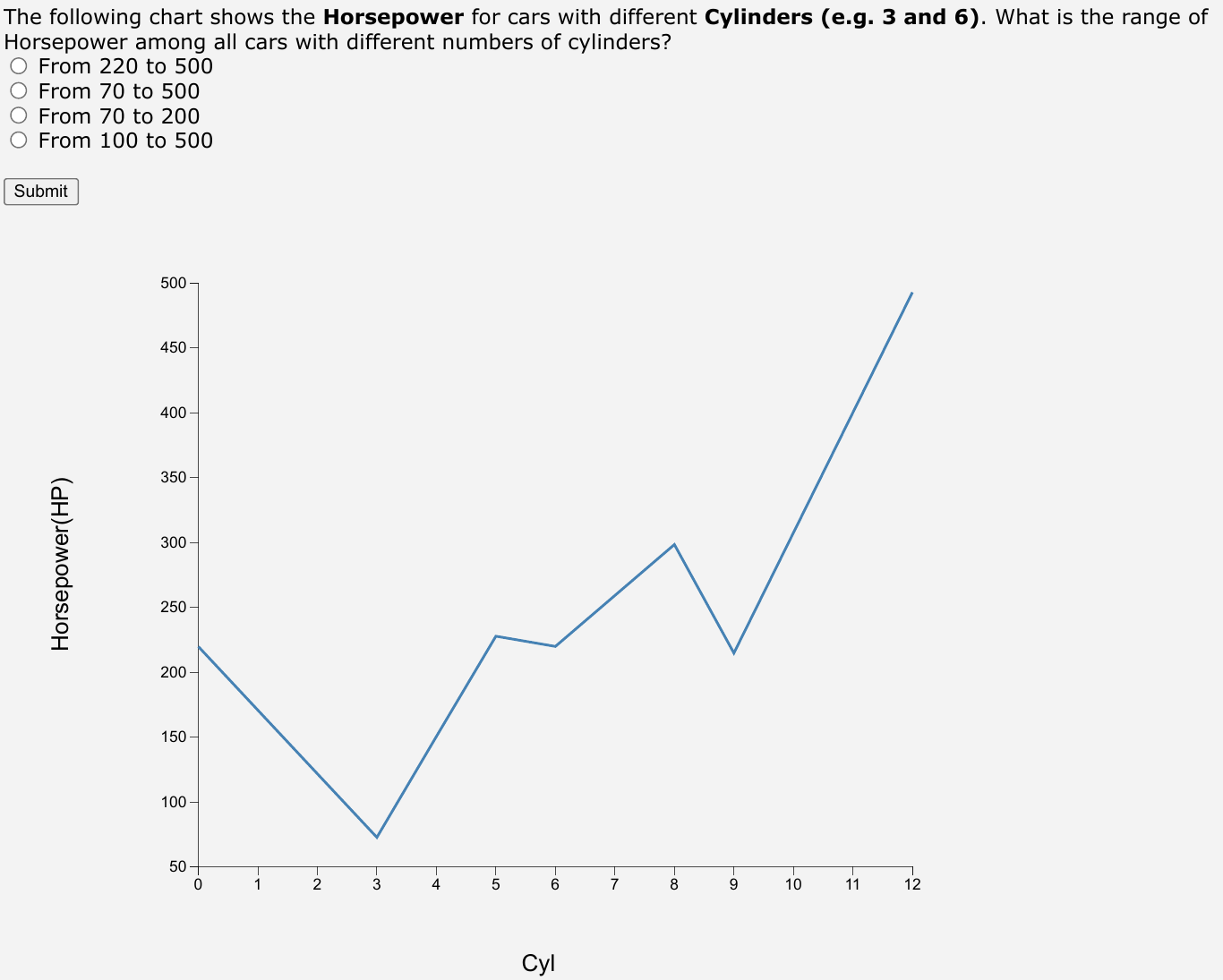}
\Description{Example questions with a line.}
}
\subfloat[\taskOr,\\\visTable]{
\includegraphics[width=0.188\textwidth]{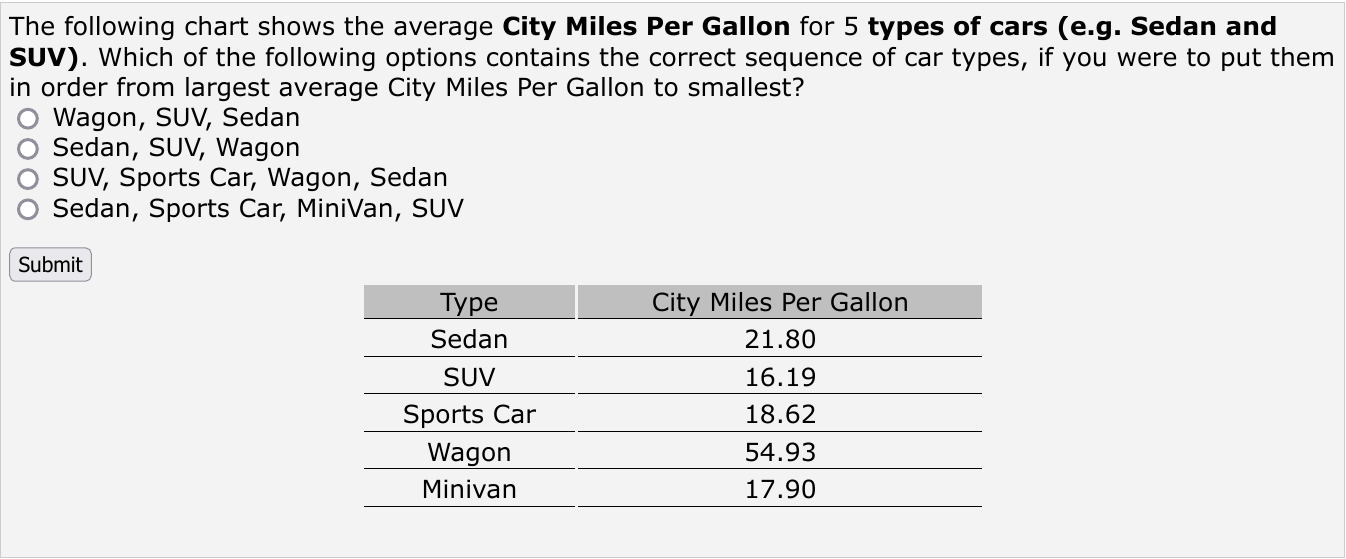}
\Description{Example questions with a table.}
}
\caption{Examples of tasks and visualizations used in the study, replicated from the prior study~\cite{saket2018task}.}
\label{fig:viz-samples}
\end{figure*}

Five small-scale (5-34 data points) two-dimensional
visualization types (\visBar, \visLine, \visPie,
\visScatter, and \visTable) were used in the study. \autoref{fig:viz-samples} shows five examples of visualizations and tasks that study participants performed, and the design of all visualizations exactly matched those of the prior study. Importantly, the study software had visualizations that were WCAG AA compliant~\cite{Developi34:online}, with WCAG AAA compliant choices of colors for all visualizations except for \visPie.  That visualization was only AA compliant due to often requiring 5 or more colors in its qualitative color palette, limiting the possibility for contrasting adjacent colors. In order to facilitate meaningful comparisons, we kept the software as-is for the PLA we recruited.

\newcommand{\exVis}{0.01em}

\section{Data Analysis and Results}
This section outlines how the study data was preprocessed (\autoref{sec:data-prep}) and modeled using Bayesian linear regression (\autoref{sec:model-desc}). It then describes the results at the task (\autoref{subsec:task-performance}) and task-visualization (\autoref{subsec:task-vis-performance}) levels. We note that, for transparency, summary results solely depicting the performance of the study participants can be found in the supplementary materials.

\subsection{Data Preparation}
\label{sec:data-prep}
To prepare the data for Bayesian modeling, we first filtered out 52 of the 200 PLA who failed the mid-study ``test'' questions, which were presented as a normal question but simply instructed the user to select \textit{True} out of the options \textit{True} and \textit{False} as per the prior study~\cite{saket2018task}. Then, to remove any unusual outliers that could bias the model (possibly resulting from conducting the study online without strict supervision), we applied a Hampel filter~\cite{pearson1999data} to all data within each age group, task, and visualization combination.  This removed a total of 151 trials (1.6\% of all trials) that had \metTime data beyond three median absolute deviations from the median, removing some noticeably large outliers we had found manually (\eDotG one trial taking over 15 minutes). While for \metTime it was easier to identify individual outlier trials, we did not do any further filtering based on \metAcc, since each trial had a binary result (correct or incorrect). 

\subsection{Modeling Description}\label{sec:model-desc}
Recent work~\cite{davis2022risks} in graphical perception has put into question the utility of design guidelines derived from strict rankings of aggregated (\eDotG average) visualization performance. Davis et al.~\cite{davis2022risks} used Bayesian modeling to illustrate that individual differences have the ability to outweigh the impacts of visualization choices. Expanding on this, we used Bayesian regression to model the interactions of task, visualization type, and age group on \metAcc and \metTime at both the population and individual levels. For each modeled performance metric, we describe the model structure and priors. 

We observe that, at first glance, the data from both our and the prior study may appear limited in their representativeness, as each participant completed trials for only one task.
This could be thought of as a dataset of ``missing'' data, where all tasks \textit{not} completed by a participant are considered missing for the purposes of further analysis. However, one strength of Bayesian models is the ability to work with and model the uncertainty of missing data~\cite{ma2018bayesian}, ultimately allowing us to generate counterfactual (simulated) participants with data for all tasks by sampling the model for posterior means.

\subsubsection{Accuracy Model}\text{ }\vspace{0.5em}

We modeled accuracy as a four-alternative forced-choice (4AFC) item test~\cite{patterson2010effects} using the \texttt{brms} package in R~\cite{burkner2017brms}, storing responses as binary values indicating a correct ($1$) or incorrect ($0$) response; examples of this data are shown in the \textit{accuracy} column of \autoref{tab:ex-data}. 

Due to its aptitude for modeling the likelihood for trials with binary outcomes~\cite{walck1996hand}, we used a Bernoulli distribution to model the accuracy for a given trial $i$, which we denoted as $A_i$. This distribution requires a single parameter $p$, which is the likelihood of a ``success'' in a trial, \iDotE answering the question correctly. Because each trial may have differing factors affecting performance, each trial $i$ required modeling a parameter $p_i$. To avoid model bias by accounting for the $25\%$ probability of randomly guessing correctly~\cite{burkner2023estimating}, we modeled $p_i$ using the equation shown in line (10) below. This approach sets the lowest possible likelihood of a correct answer as $25\%$, modeling the ``remaining'' possible $75\%$ as $\tfrac{3}{4}\cdot\text{logit}^{-1}(\eta_i)$. 
In line with Davis et al.~\cite{davis2022risks}, we used a mean submodel that we denoted as $\eta_i$ which captured both population- and participant-level information for estimating $p_i$. The population-level information, $\beta_{\text{vis}[i],\text{task}[i],\text{age}[i]}$, models the average accuracy across all trials with the given combination of visualization, task, and age group present in trial $i$. To make sure that the model internalizes individual differences (\iDotE some participants may perform better or worse with certain visualizations), we then added a participant-level component, $U_{\text{vis}[i],\text{part}[i]}$, that adjusts the mean based on an individual participant's performance with the given visualization.

\small
\begin{align*}
    &V = 5&\text{\# of types of visualizations}\\
    &K = 6&\text{\# of questions per visualization}\\
    &P = 328&\text{\# of YA (180) \& PLA (148)}\\
    &i \in \{1, 2, \ldots, VKP\}&\text{trial index}\\
    &D= 10 &\text{\# of tasks}\\
    &\text{vis}[i] \in \{1,\ldots,V\}&\text{vis. used in trial $i$}\\
    &\text{task}[i] \in \{1,\ldots,D\}&\text{task for trial $i$}\\
    &\text{age}[i] \in \{1,2\}&\text{age group for trial $i$}\\
    &\text{part}[i] \in \{1, \ldots, P\}&\text{participant for trial $i$}\\[15pt]
    &A_i \sim \text{Bernoulli}(p_i)&\text{likelihood for trial $i$}\\
    &p_i = \tfrac{1}{4} + \tfrac{3}{4}\cdot\text{logit}^{-1}(\eta_i)&\text{likelihood}\\
    &\eta_i = \beta_{\text{vis}[i],\text{task}[i],\text{age}[i]} + U_{\text{vis}[i],\text{part}[i]}&\text{mean submodel}
\end{align*}
\vspace{0.5em}

\normalsize

In our case, we had at most $VKP = 5\cdot6\cdot328=9840$ trials in our combined data set of YA from the prior study and PLA we collected, however, due to filtering (described in \autoref{sec:data-prep}), the final number was slightly smaller. 
Crucially, our $\beta$ term modeled the \textit{interactions} of the three categorical variables, \iDotE we aimed for the model to internalize how each of them interrelate and influence participants' accuracy. To illustrate this, our $\eta_i$ term would be written in Wilkinson-Pinheiro-Bates syntax~\cite{wilkinson1973symbolic,pinheiro2017package} in \texttt{brms} as:
\begin{center}
\vspace{0.5em}
\small\texttt{eta $\sim$ (vis * task * age.group) + 0 + (vis + 0 | pt | user.id)}
\vspace{0.5em}
\end{center}
In that syntax, $\beta_{\text{vis}[i],\text{task}[i],\text{age}[i]}$ corresponds to \texttt{(vis * task * age.group) + 0}, with the asterisks indicating that we are modeling the \textit{interactions} between the three categories as opposed to independent impacts on performance. Meanwhile, $U_{\text{vis}[i],\text{part}[i]}$ corresponds to \texttt{(vis + 0 | pt | user.id)} to indicate that it is modeling random effects based on the individual participant referenced by their unique identifier. In line with the approach of Davis et al., we used the~\texttt{+~0} syntax to one-hot code each categorical variable to guarantee unique coefficients, while the~\texttt{|~pt~|} syntax shared the covariance matrix of the random effects across all submodels~\cite{davis2022risks}.
We especially diverged from Davis et al.~\cite{davis2022risks} by having multiple categorical variables impacting the resulting likelihood; moreover, the ways that varying combinations of visualization, task, and age conditionally affected the mean accuracy necessitated not treating them as independent~\cite{mcelreath2018statistical}. Because each participant experienced all five visualizations but for only one task and as a member of only one age group, we only modeled random effects at the participant-level, $U_{\text{vis}[i],\text{part}[i]}$, based on the visualization. 

\begin{table}[ht]
\caption{Example trial-level data. Columns are the data type (\eDotG \textbf{user.id} is a unique identifier for a participant), and each row is a single data point (trial) storing their identifier, the visualization (\textbf{vis}) shown, the \textbf{task} (\iDotE the type of question asked), their answer's \textbf{accuracy} (1 if correct, 0 if incorrect), the \textbf{time} (s) taken to answer the question, and the participant's age group (\textbf{age.group}).
}
\label{tab:ex-data}
\resizebox{\columnwidth}{!}{%
\begin{tabular}{llllll}
\hline
\multicolumn{1}{c}{\textbf{user.id}} & \multicolumn{1}{c}{\textbf{vis}} & \multicolumn{1}{c}{\textbf{task}} & \multicolumn{1}{c}{\textbf{accuracy}} & \multicolumn{1}{c}{\textbf{time}} & \multicolumn{1}{c}{\textbf{age.group}} \\ \hline
ya1 & Table & Anomalies & 1 & 9 & YA \\
ya1 & PieChart & Anomalies & 0 & 24 & YA \\
... & ... & ... & ... & ... & ... \\
ya2 & BarChart & Extremum & 1 & 10 & YA \\
... & ... & ... & ... & ... & ... \\
pla1 & Scatterplot & Order & 1 & 13 & PLA \\ \hline
\end{tabular}%
}
\end{table}

Due in part to a lack of existing benchmarks for this type of data, we used a weakly-informed prior~\cite{lemoine2019moving} for our mean submodel, consistent with other studies in HCI where pre-existing data was limited~\cite{davis2022risks,sarma2024odds,kay2016researcher}. This prior roughly covers the range of expected possible values for mean accuracy, lightly nudging the posterior distribution toward more likely values but ultimately allowing the likelihood to have a more dominant influence on the posterior distribution due to the large amount of data collected~\cite{gelman2019prior}.
We employed a standard normal prior for $\eta_i$, \iDotE $\eta_i \sim \text{Normal}(0,1)$, resulting in a 95\% confidence that mean accuracies would fall in a $[0.34, 0.91]$ range while 99\% would fall in a $[0.29, 0.96]$ range. 
We believed that most participants would get at least 25\% correct due to the multiple choice format, but with 30 total questions it could be difficult to get all of them correct. 
For all other priors, we used the default weakly-informed priors provided by \texttt{brms}. 
The model ran with a total of six chains and $12000$ post-warmup draws, achieving bulk effective sample sizes in the $[1290, 7780]$ range for participant-level effects~($U$) and in the $[1280, 6240]$ range for population-level effects~($\beta$), indicating stable estimates~\cite{burkner2017brms}. All effects had $\hat{R} \leq 1.01$, showing proper model convergence to a stable solution~\cite{vehtari2021rank}. 
We then ran a sensitivity analysis by fitting additional models with small perturbations to the $\eta_i$ prior and comparing them using interval analysis~\cite{cumming2014new}, finding no evidence of differences in those models' regression coefficients and corroborating our prior's robustness.

After fitting the model to the participant data, we sampled from the posterior distribution to generate \textit{counterfactual} participants, which have data regarding their performance across all task and visualization combinations~\cite{mcelreath2018statistical}. This allowed us to leverage the data we collected to expand the size of our sample, ultimately making our analyses more robust and generalizable. 
We thus sampled the mean accuracy of 12000 counterfactual users, each for every possible combination of age (2), task (10), and visualization (5), resulting in a total of $12000 \times (2 \times 10 \times 5)=$ 1.2 million posterior means. 

\subsubsection{Time Model}\text{ }\vspace{0.5em}

We modeled time at the per-trial level as positive, continuous data using a lognormal distribution~\cite{mcelreath2018statistical}, which is well-suited for right-skewed data such as response times~\cite{rummel2014probability, dragicevic2016fair}. This distribution has two parameters, a location $\mu$ and scale $\sigma$. Thus, the time $T_i$ for a trial $i$ can be modeled as a lognormal distribution with location $\mu_i$ and scale $\sigma_i$. Similar to the accuracy model, we modeled population-level patterns of performance with both $\beta^{(\mu_i)}$ and $\beta^{(\sigma_i)}$ and participant-level information with $U^{(\mu_i)}$ and $U^{(\sigma_i)}$, noting that each distribution parameter ($\mu, \sigma$) required its own submodel.

\small
\begin{align*}
    &V = 5&\text{\# of visualizations}\\
    &K = 6&\text{\# of questions per visualization}\\
    &P = 328&\text{\# of YA (180) and PLA (148)}\\
    &i \in \{1, 2, \ldots, VKP\}&\text{trial index}\\
    &D= 10 &\text{\# of tasks}\\
    &\text{vis}[i] \in \{1,\ldots,V\}&\text{visualization for trial $i$}\\
    &\text{task}[i] \in \{1,\ldots,10\}&\text{task for trial $i$}\\
    &\text{age}[i] \in \{1,2\}&\text{age group for trial $i$}\\
    &\text{part}[i] \in \{1, \ldots, P\}&\text{participant for trial $i$}\\[15pt]
    &T_i \sim \text{LogNormal}(\mu_i, \sigma_i)&\text{likelihood, observation $i$}\\
    &\mu_i = \beta^{(\mu_i)}_{\text{vis}[i],\text{task}[i],\text{age}[i]} + U^{(\mu_i)}_{\text{vis}[i],\text{part}[i]}&\text{location submodel}\\
    &\sigma_i = \beta^{(\sigma_i)}_{\text{vis}[i],\text{task}[i],\text{age}[i]} + U^{(\sigma_i)}_{\text{vis}[i],\text{part}[i]}&\text{scale submodel}
\end{align*}
\vspace{0.5em}

\normalsize

In this case, our formulation yielded the following in \texttt{brms} syntax, where the \texttt{*} character indicates interactions between variables: 
\begin{center}
\vspace{0.5em}
\small\texttt{time $\sim$ (vis * task * age.group) + 0 + (vis + 0 |pt| user.id)}
\vspace{0.5em}
\end{center} 
For the location and scale we modeled random effects at the participant level using $U_{\text{vis}[i],\text{part}[i]}$, additionally using the~\texttt{+ 0} to one-hot code our categorical variables and the~\texttt{| pt |} syntax to share covariance matrices of the random effects across submodels.

Our weakly-informed prior~\cite{lemoine2019moving} for the location submodel was a normal distribution $\mu_i \sim \text{Normal(3.4,1)}$ which, in log space, has a 95\% chance of covering mean response times in the $[4,221]$ second range and a 99.7\% chance of covering the $[1.5, 602]$ second range. This range of times covers minimums that we found reasonable for actually answering the question as well as maximums that, although long, could reasonably be expected (in rare occasions) for answering a question. We used the built-in weakly-informed priors from \texttt{brms} for all other priors, including the variance. The model ran with 6 chains and 12000 post-warm up draws, with bulk effective sample sizes for participant-level effects ($U$) in the $[1109,5717]$ range and population level effects in the $[1544,4950]$ range, showing model stability~\cite{burkner2017brms}. All effects had $\hat{R} \leq 1.01$, showing convergence to a stable solution~\cite{vehtari2021rank}. Lastly, we sampled $12000$ counterfactual users to obtain $12000 \times (2 \times 10 \times 5)=$ 1.2 million posterior means for further analysis. Similar to accuracy, we ran a sensitivity analysis by fitting additional models with small perturbations to the $\mu_i$ prior, finding no evidence of differences in the models' regression coefficients and further showing the reliability of our weakly-informed prior.

\subsection{Task Performance}\label{subsec:task-performance}
Prior research shows that the performance of people varies across data analysis tasks~\cite{quadri2021survey}. There are tasks that people (averaged across individuals) appear to find more time-consuming and prone to errors than others and vice versa. For instance, tasks like \taskCo and \taskAn tend to be more prone to errors, and tasks such as \taskRe and \taskEx are typically faster to perform~\cite{saket2018task, amar2005low, scaife1996external, heer2010crowdsourcing}. This variation raises a critical question about the impact of aging on these dynamics. Specifically, we sought to explore whether the analysis tasks that PLA find particularly time-consuming or difficultly-intricate notably differ from those of YA. 
However, this analysis notably not only aims to ascertain whether there are differences in overall performance (comparing averages), but also whether there are differences in the spread of individual performances (comparing distributions). 

\begin{figure*}[ht]
  \centering
      \includegraphics[width=\linewidth]{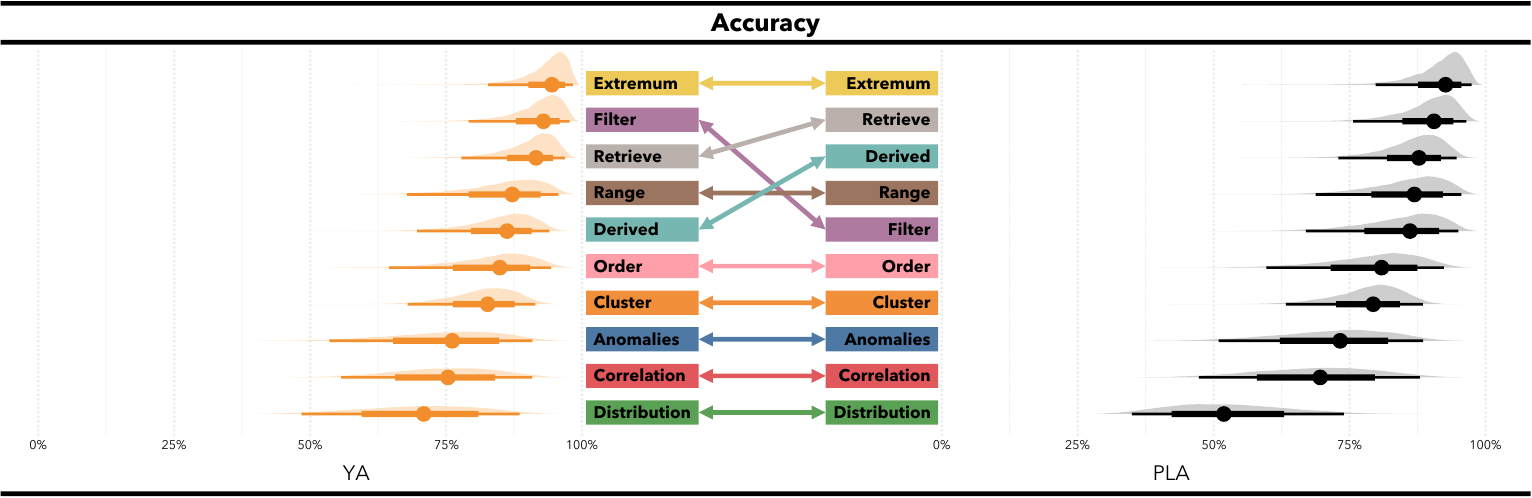}
  \caption{This figure shows the distribution of mean \metAcc, aggregated by task. Data shown is the distribution as well as the 50\textsuperscript{th} (median), 66\textsuperscript{th}, and 95\textsuperscript{th} percentiles of all mean accuracies for the 12000 counterfactual participants, aggregated across all tasks for YA~\yacbox and PLA~\placbox. Tasks are color-coded and sorted highest-to-lowest by median value, and arrows connect tasks across age groups.}\label{fig:task-accuracy}
  \Description{Plot of accuracy for YA and PLA across all visualizations. The most noticeable differences in rank are for Filter and Derived.}
\end{figure*}

We began the analysis by calculating the spread of YA and PLA's \textit{per-task} performance for both \metAcc and \metTime metrics by using the sampled posterior means described in \autoref{sec:model-desc}. This entailed aggregating the means across all visualizations for a given task (\eDotG~\taskDi), performance metric (\eDotG \metAcc), and age category (\eDotG PLA) into a single vector and then calculating their 50\textsuperscript{th} (median), 66\textsuperscript{th},  and 95\textsuperscript{th} percentiles. Grounded in these analyses, we proceeded to rank analysis tasks by median according to the performance metrics for each of the age groups. \autoref{fig:task-accuracy} and \autoref{fig:task-time} present performance in aggregate, showing the distribution of each age group's mean per-task \metAcc and \metTime, respectively. Meanwhile, \autoref{fig:task-acc-rank-pct} and \autoref{fig:task-time-rank-pct} illustrate performance more focused at the individual level, showing the proportions of each age group that would achieve their best, middle, and worst performance with various subsets of tasks. To compare performance for each metric between the two age groups for a given task we use the \textit{percent difference} of the groups' medians, $m_Y$ for YA and $m_P$ for PLA, calculated as $\frac{|m_Y-m_P|}{\frac{1}{2}(m_Y+m_P)}$; these results are provided in \autoref{tbl:pct-diff}. This approach is order-agnostic to its inputs and stays scaled to the data's original units, while also allowing for more direct comparison between two different metrics (\iDotE \metAcc and \metTime), thus making it well-suited for comparison~\cite{dragicevic2012my}.

\subsubsection{Accuracy}\text{ }\vspace{0.5em}

\textbf{Aggregate Level.} 
We found that both YA and PLA achieved their highest accuracy when performing the task~\taskEx and lowest accuracy with the task~\taskDi.
Aside from the task~\taskDi, which showed a 31\% difference in \metAcc between the two groups, the accuracy for most tasks was comparable, resulting in relatively low percentage differences.
Furthermore, the relative rankings of tasks were closely aligned between YA and PLA, with the largest differences occurring for \taskFi and \taskDe. 
We next examined the correlation between the \metAcc-driven task rankings using Kendall's tau~\cite{noether1981kendall}. We considered a higher positive correlation to indicate a greater alignment between the two groups and vice versa. As a result, we found a Kendall rank correlation coefficient of 0.822~$(p < 0.005)$, indicating a strong positive relationship between the two age groups' rankings of tasks. This strong positive correlation implies a considerable degree of alignment in the experience of task difficulty (in terms of \metAcc) between YA and PLA.

\begin{figure*}[ht]
  \centering
      \includegraphics[width=\linewidth]{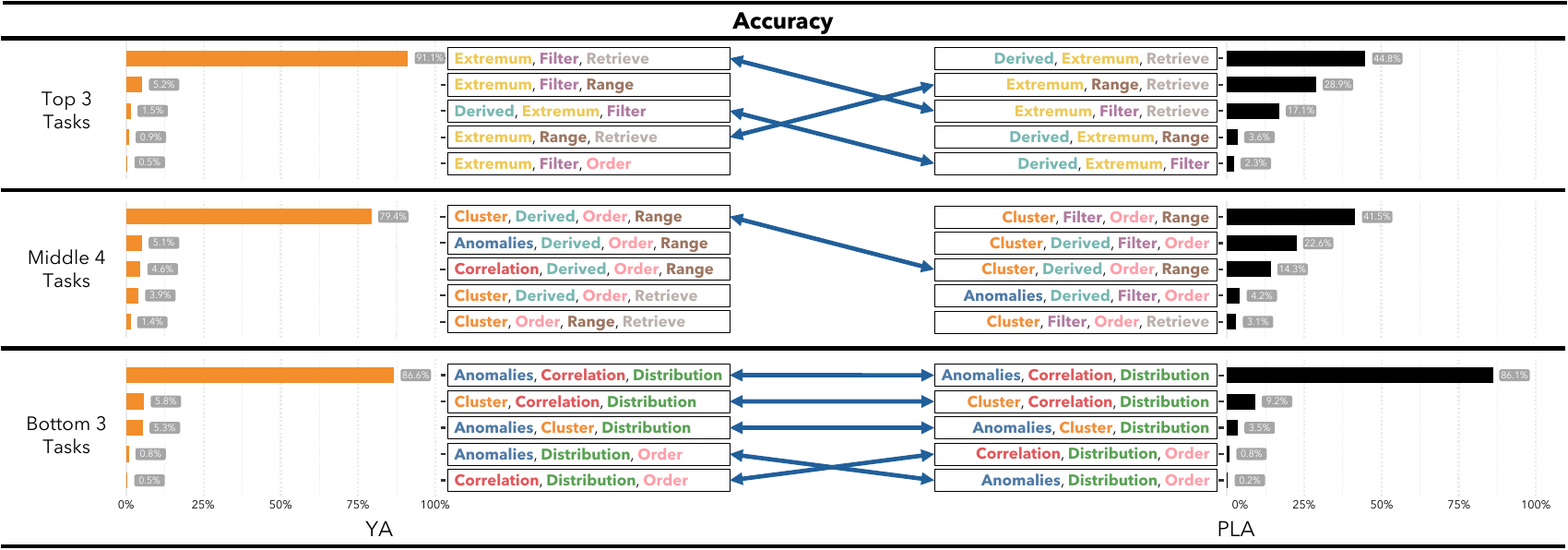}
  \caption{
  This figure depicts which sets of tasks were most commonly the three highest (Top 3), four middle (Middle 4), and 3 lowest (Bottom 3) rankings for \metAcc across all visualizations for both YA~\yacbox and PLA~\placbox. Sets are presented in alphabetical order, \iDotE YA may have had different rankings between \taskEx, \taskFi, and \taskRe, however overall $91.1\%$ of counterfactual participants had those tasks as their three highest-accuracy tasks. Arrows between sets from YA to PLA indicate a shared set of tasks.}\label{fig:task-acc-rank-pct}
  \Description{Proportions of simulated participants who had sets of tasks as their top 3 (best), middle 4, and bottom 3 (worst) tasks for accuracy.}
\end{figure*}
\textbf{Individual Level.}
With recent work emphasizing the importance of also understanding performance at the individual level~\cite{davis2022risks}, we further aimed to ascertain how often certain task orderings occurred per-person. However, to keep analysis reasonable and succinct with the number of possible orderings, we decided to count how often counterfactual participants found sets of tasks as their three top-, four middle-, and three worst-performing tasks, hereafter referred to as \textit{partitions}. This allowed us further understand which tasks each age group found more or less difficult as well as how those tasks were distributed among each age group, with results shown in \autoref{fig:task-acc-rank-pct}.
We observed that each partition for YA had one predominant set of tasks, while PLA had more diversity in their top three and middle four. However, there was strong agreement within each group for the task in their bottom three partitions.

Between the two age groups, the top partition shared three of five sets (though with differing ranks and amounts), the middle four only shared the set \{\taskCl, \taskDe, \taskOr, \taskRa\}, and the bottom three had similar distributions \textit{and} sets of tasks, including the same set \{\taskAn, \taskCo, \taskDi\} as the most common. The main observed difference between the two age groups was that \taskDe was more common in the top partition for PLA than YA, whereas \taskFi was more common in the top partition for YA than PLA; this pattern aligns with the results of the aggregate analysis, where those two tasks also had the largest differences in rank. These results indicate that PLA may experience differing amounts of difficulty with some tasks compared to YA and with greater heterogeneity within their age group. 
Furthermore, the similarity in task sets and proportions between both of the age groups for the three worst-performing tasks indicate that the tasks found ``most difficult'' appear to not noticeably change with age.

\begin{figure*}[ht]
    \centering
    \captionsetup{type=table}
    \captionof{table}{Percent Differences between YA and PLA in median (a) \metAcc and (b) \metTime, per task. A larger percent difference indicates greater performance differences between the two groups for that task.}
    \label{tbl:pct-diff}
    \begin{subfigure}{\columnwidth}
        \centering
        \caption{Accuracy}
        \begin{tabular}{rr}
        \hline
        \textbf{Task} & \textbf{\% Difference} \\ \hline
        \taskAn & 3.8\% \\
        \taskCl & 4.1\% \\
        \taskCo & 7.9\% \\
        \taskDe & 1.7\% \\
        \taskDi & 31.0\% \\
        \taskEx & 1.9\% \\
        \taskFi & 7.6\% \\
        \taskOr & 4.8\% \\
        \taskRa & 0.3\% \\
        \taskRe & 1.1\% \\ \hline
        \end{tabular}
        \label{tbl:pct-diff-acc}
        \vspace{2em}
    \end{subfigure}
    \begin{subfigure}{\columnwidth}
        \centering
        \caption{Time}
        \begin{tabular}{rr}
        \hline
        \textbf{Task} & \textbf{\% Difference} \\ \hline
        \taskAn & 94.2\% \\
        \taskCl & 49.3\% \\
        \taskCo & 68.8\% \\
        \taskDe & 48.9\% \\
        \taskDi & 36.0\% \\
        \taskEx & 41.2\% \\
        \taskFi & 79.4\% \\
        \taskOr & 48.2\% \\
        \taskRa & 37.3\% \\
        \taskRe & 61.4\% \\ \hline
        \end{tabular}
        \label{tbl:pct-diff-time}
    \end{subfigure}
\end{figure*}

\subsubsection{Time}\text{ }\vspace{0.5em}

\begin{figure*}[ht]
  \centering
      \includegraphics[width=\linewidth]{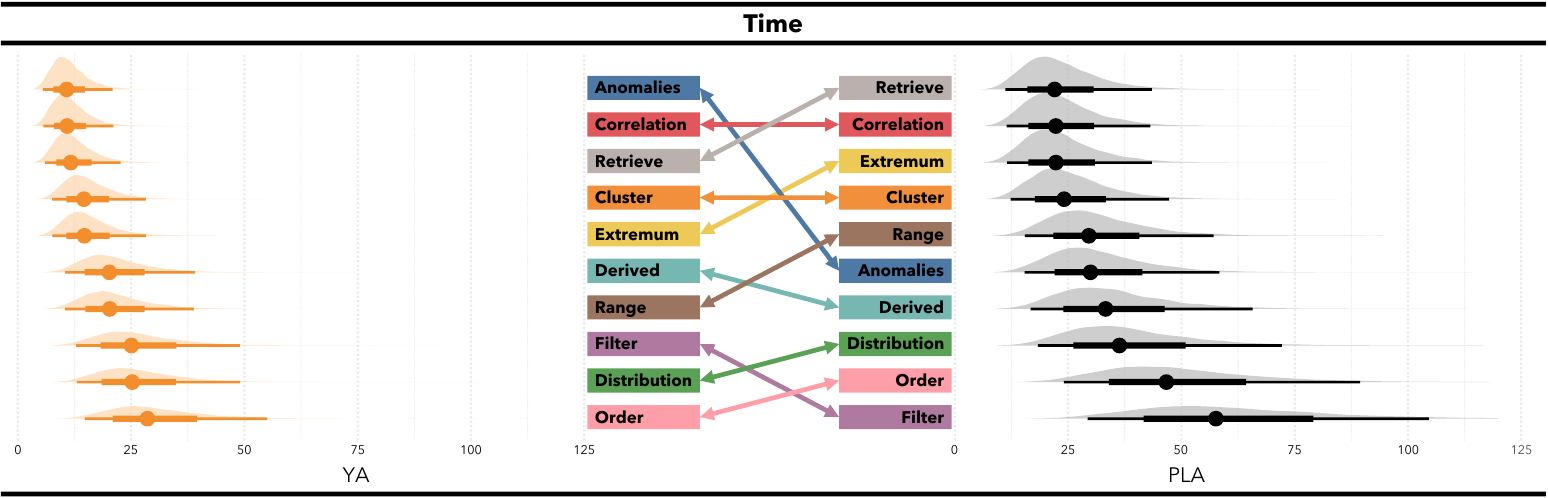}
  \caption{This figure shows the distribution of mean \metTime (s), aggregated by task. Data shown is the distribution as well as the 50\textsuperscript{th} (median), 66\textsuperscript{th}, and 95\textsuperscript{th} percentiles of all mean times for the 12000 counterfactual participants, aggregated across all tasks for YA~\yacbox and PLA~\placbox. Tasks are color-coded and sorted lowest-to-highest by median value, and arrows connect tasks across age groups.}\label{fig:task-time}
  \Description{Plot of time for YA and PLA across all visualizations. The most noticeable differences in rank are for Anomalies, Retrieve, Extremum, Range, and Filter.}
\end{figure*}
\textbf{Aggregate Level.}
As shown in \autoref{fig:task-time}, we observed that YA were fastest completing \taskAn, \taskCo, and \taskRe while needing the most time for \taskOr; meanwhile, PLA were fastest completing \taskRe, \taskCo, and \taskEx while taking the most time to complete \taskFi. In comparison to \metAcc, we observed much greater differences in time taken between the age groups across all tasks, with the highest percent difference for \metAcc being less than the lowest percent difference for \metTime. Interestingly, this means that YA and PLA were both most different in \metAcc and most similar in time needed for performing \taskDi. The largest time differences between groups occurred for \taskAn and \taskFi, which were among the largest differences in task ranking; \taskRe, \taskRa, and \taskEx showed the next-highest differences in task ranking. Compared to \metAcc, we observed more small differences in ranks between groups, with only \taskCo and \taskCl sharing ranks. Thus, as was done for \metAcc, we then used Kendall's tau to calculate the correlation between the two age groups' time-driven task rankings. We found a Kendall rank correlation coefficient of 0.556 ($p < 0.05$), indicating a moderate positive relationship between the two age groups, though not as strong as for \metAcc. 
We also want to note the wider value ranges for the 66\textsuperscript{th} and 95\textsuperscript{th} percentiles for PLA, indicating greater diversity in completion time at the task level.

\begin{figure*}[ht]
  \centering
      \includegraphics[width=\linewidth]{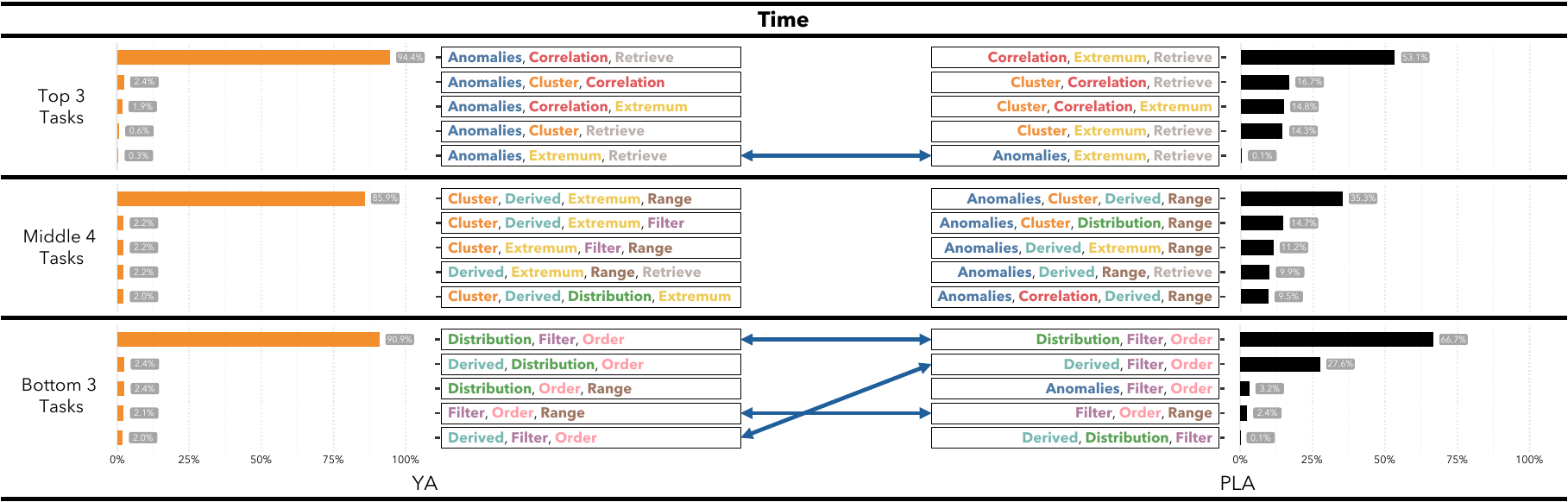}
  \caption{This figure depicts which sets of tasks were most commonly the three highest (Top 3), four middle (Middle 4), and 3 lowest (Bottom 3) rankings for \metAcc across all visualizations for both YA~\yacbox and PLA~\placbox. Sets are presented in alphabetical order, \iDotE PLA may have had different rankings between \taskDi, \taskFi, and \taskOr, however overall $66.7\%$ of counterfactual participants had those tasks as their three slowest tasks. Arrows between sets from YA to PLA indicate a shared set of tasks.}\label{fig:task-time-rank-pct}
  \Description{Proportions of simulated participants who had sets of tasks as their top 3 (best), middle 4, and bottom 3 (worst) tasks for time.}
\end{figure*}

\textbf{Individual Level.}
As with \metAcc, we counted how often counterfactual participants had sets of tasks as their top, middle, and bottom performances, shown in \autoref{fig:task-time-rank-pct}. 
For each partition, there was strong agreement within the YA population. Meanwhile, PLA had a greater spread in their top and middle partitions. Although not as much as accuracy, PLA still showed the strongest agreement for the bottom partition compared to both the middle and top.

Compared to \metAcc, there was much less overlap in sets of tasks in the most common top three and middle four, with only \{\taskAn, \taskEx, \taskRe\} shared but both at low percentages.  This is likely due to \taskAn appearing more often in the top three for YA and middle four for PLA, except for their aforementioned overlap. The bottom partition shares three of five sets, including the most-common set for both age groups \{\taskDi, \taskFi, \taskOr\}. These results indicate some noticeable differences in how long tasks take for YA and PLA, however the most time-consuming tasks appear to roughly stay the same with age. Furthermore, PLA exhibited much greater heterogeneity in the distribution of tasks requiring the least and middle amounts of time, consistent with the results for \metAcc.

\subsection{Task-Visualization Performance}\label{subsec:task-vis-performance}
The outcomes of the first stage of our analysis suggested that, overall, YA and PLA appear to find tasks similarly difficult and time-consuming relative to each other. 
However, that analysis did not account for the influence of visualization choices on the counterfactual participants' performance across the various tasks.
Hence, we next took a deeper look at YA and PLA's \metAcc and time for each analysis task, further broken down by visualization type. Similar to \autoref{subsec:task-performance}, we aimed to understand if any differences exist at a more aggregated level as well as at the more individual level.

We first analyzed the distribution of each age group's performance for \metAcc and \metTime by aggregating the model's sampled posterior means across a given task (\eDotG~\taskAn), visualization (\eDotG~\visBar), performance metric (\eDotG~\metAcc), and age group (\eDotG YA) into a single vector and computed their 50\textsuperscript{th} (median), 66\textsuperscript{th}, and 95\textsuperscript{th} percentiles.  For each task and each age group, we then sorted the visualizations by median value, as shown in \autoref{fig:task-chart-accuracy} and \autoref{fig:task-chart-time} for \metAcc and \metTime, respectively. For greater insights into performance at the individual level, we used the sample means to count the proportions of counterfactual participants 
who experienced each visualization at each rank (1-5) per task,
as shown in \autoref{fig:rank-dist-acc} (\metAcc) and \autoref{fig:rank-dist-time} (\metTime). Percent differences between the groups' medians for each task and metric were also calculated for comparison, which are provided in \autoref{tbl:pct-diff-tv}.

\subsubsection{Accuracy}\text{ }\vspace{0.5em}

\begin{figure*}[ht]
    \centering
    \begin{subfigure}[t]{0.475\textwidth}
        \centering
        \includegraphics[width=\linewidth]{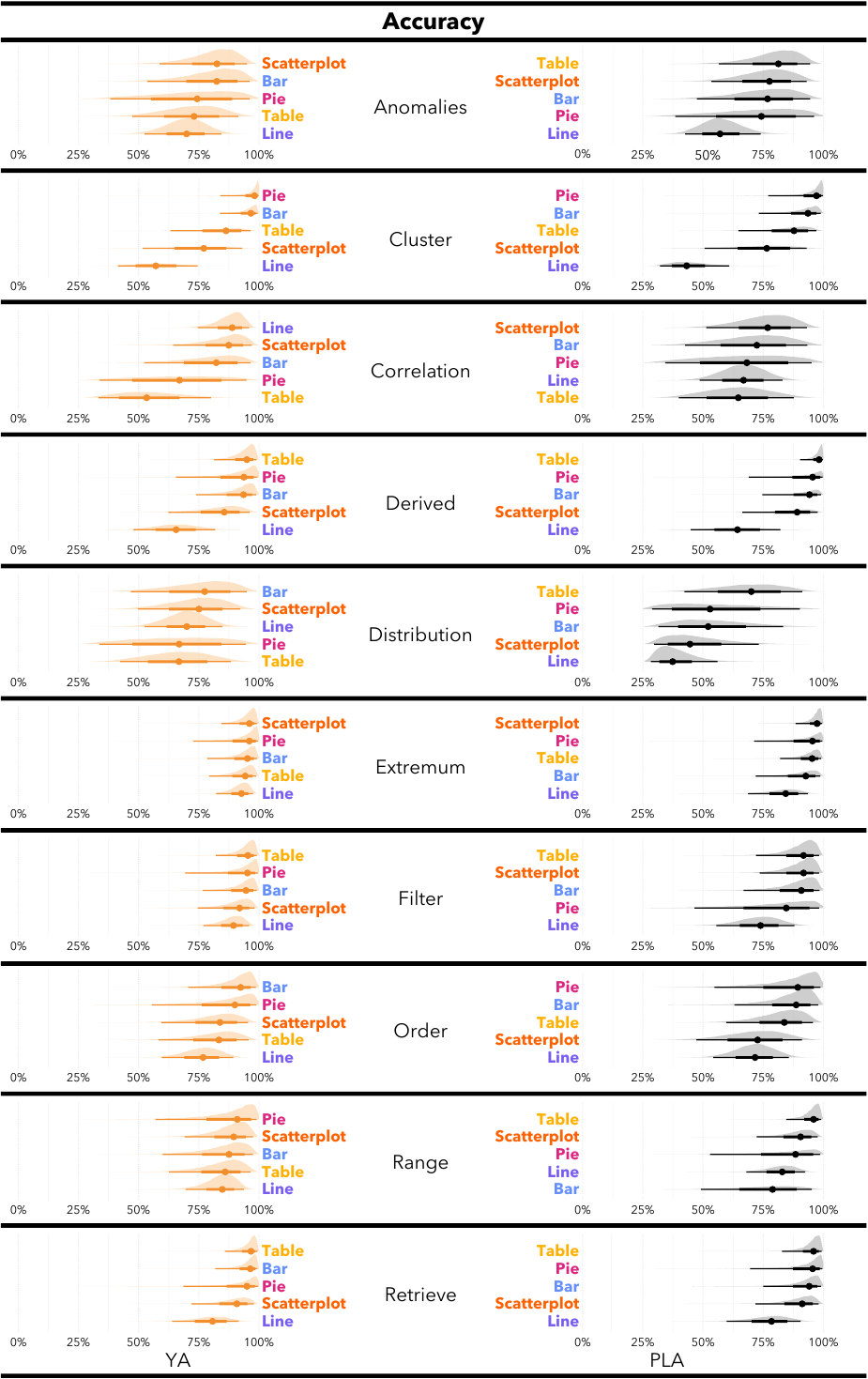}
        \caption{This figure depicts the distribution of mean \metAcc for counterfactual YA~\yacbox (left) and PLA~\placbox (right), aggregated by task and visualization. Visualizations are sorted by median from best to worst, and each distribution also shows the 50\textsuperscript{th}, 66\textsuperscript{th}, and 95\textsuperscript{th} percentiles.}\label{fig:task-chart-accuracy}
        \Description{Accuracy plots for YA and PLA, per visualization and task. Values are pretty similar, but Table is ranked higher for people in late adulthood more often as the top-ranked visualization.}
    \end{subfigure}
    \hfill
    \begin{subfigure}[t]{0.485\textwidth}
        \centering
        \includegraphics[width=\textwidth]{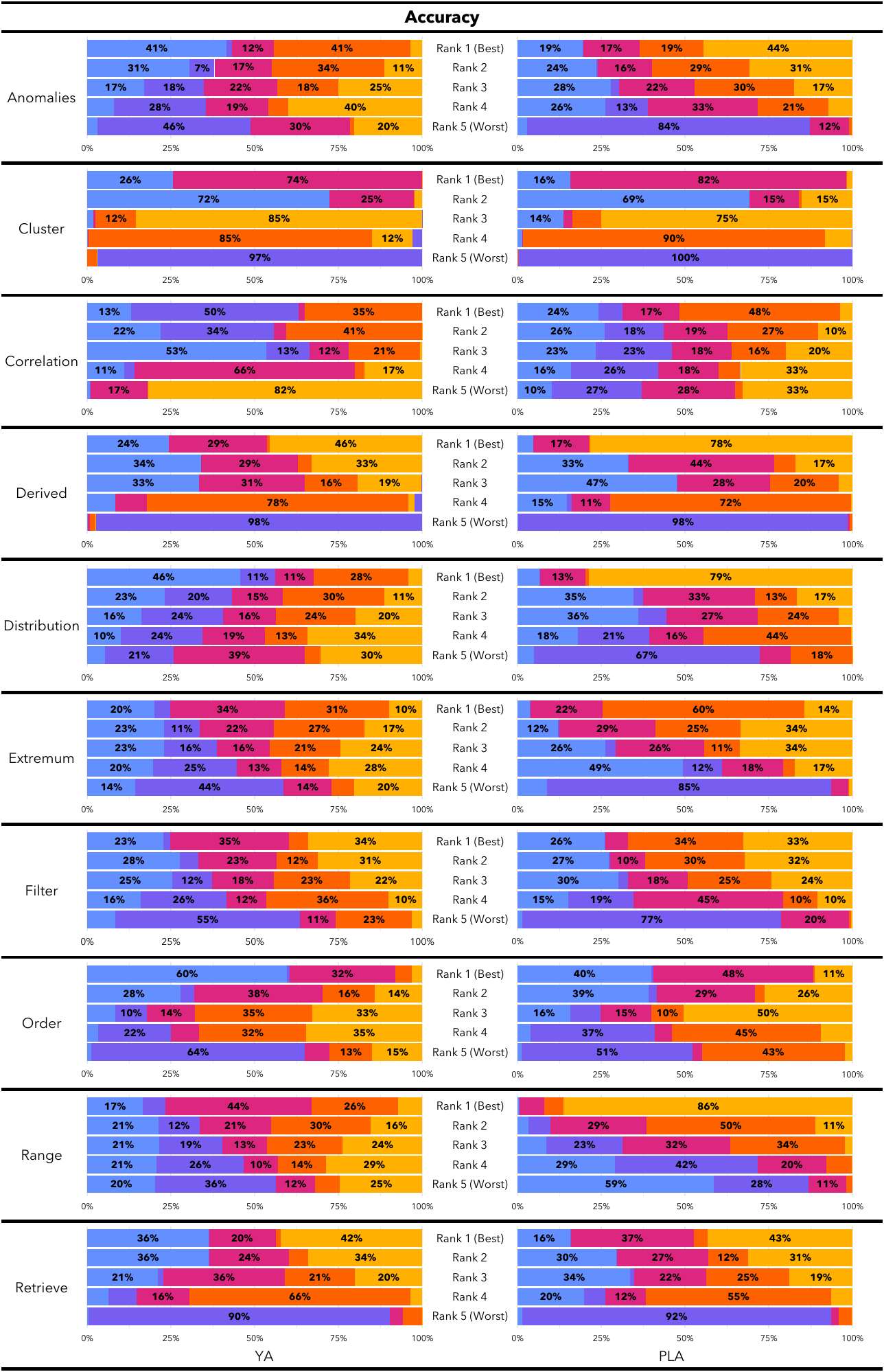}
        \caption{This figure shows the proportions of counterfactual participants with each visualization (\visBar~\barcbox, \visLine~\linecbox, \visPie~\piecbox, \visScatter~\scattercbox, and \visTable~\tablecbox) at each rank for accuracy for a given task. YA are depicted on the left, PLA on the right. 
        Any bars without a text label are less than 10\%. 
        }\label{fig:rank-dist-acc}
        \Description{Proportions of simulated participants who had each visualization ranked 1-5 for each task for accuracy.}
    \end{subfigure}
    \caption{Task-visualization accuracy results at the (a) aggregate and (b) individual level.}
\end{figure*}

\textbf{Aggregate Level.}
Provided in \autoref{fig:task-chart-accuracy}, there was a variety of visualizations leading to highest \metAcc.  For YA \visBar was best for \taskDi and \taskOr, \visLine was best for \taskCo, \visPie was best for \taskCl and \taskRa, \visScatter was best for \taskAn and \taskEx, and \visTable was best for \taskDe, \taskFi, and \taskRe.  Meanwhile, for PLA \visBar and \visLine were not top-ranked for any tasks, \visPie was best for \taskCl and \taskOr, \visScatter was best for \taskCo and \taskEx, and \visTable was best for \taskAn, \taskDe, \taskDi, \taskFi, \taskRa, and \taskRe. We also observed noticeably lower \metAcc for both groups using \visLine compared to the other visualizations for the \taskCl, \taskDe, and \taskRe tasks; furthermore, YA had notably lower accuracy using \visTable for \taskCo, while PLA also experienced notably lower accuracy with \visLine for \taskAn, \taskEx, and \taskFi.

YA and PLA shared some agreement on the best and worst visualizations for task performance.  For \taskCl, \taskDe, \taskEx, \taskFi, and \taskRe, the two age groups achieved their highest accuracy with the same visualization.  Meanwhile, YA and PLA, respectively, had differing best visualizations for \taskAn, \taskCo, \taskDi, \taskOr, and \taskRa. For worst-performing visualizations, the groups only differed for \taskDi. The percent differences between the two groups in accuracy with their best-performing visualizations (shown in \autoref{tbl:pct-diff-tv-time}) were mostly minor ($< 6\%$ difference), with the largest being \taskCo and \taskDi. The largest percent differences between age groups for the same task-visualization pair occurred for \{\taskDi, \visLine\},  \{\taskDi, \visScatter\}, and \{\taskDi, \visBar\}. However, there were other noteworthy disagreements on visualization choice, such as \visLine for \taskCo, \visTable for \taskDi, \visTable for \taskAn, and \visTable for \taskRa. With all this in mind, it appears that PLA, overall, achieve task accuracy similar to YA, however in some cases they may do so with a different type of visualization. Additionally, PLA appeared to excel using \visTable more often than YA. 

\textbf{Individual Level.}
\autoref{fig:rank-dist-acc} displays the distributions of counterfactual participants that achieved each rank with each visualization, per task, which can allow us to understand whether one visualization was predominantly the ``best'' or ``worst'' for a given task and age group, or if instead there was more parity among the visualization options. YA had a majority ($> 50\%$) ``best'' visualization for \taskCl, \taskCo, and \taskOr, while PLA had a majority for \taskCl, \taskDe, \taskEx, and \taskRa. Meanwhile, YA had \visLine as the majority ``worst'' visualization for \taskCl, \taskDe, \taskFi, \taskOr, and \taskRe, with \taskCo instead doing so with \visTable; PLA had \visLine as a majority ``worst'' for \taskAn, \taskCl, \taskDe, \taskDi, \taskEx, \taskFi, \taskOr, and \taskRe, with \taskRa instead doing so with \visBar. Interestingly, YA had four tasks (\taskAn, \taskDi \taskEx, and \taskRa) that had no majority visualization for any rank whereas PLA only had no majority across ranks for \taskCo, possibly indicating slightly greater heterogeneity in YA at the task-visualization level.

The tasks with the most-similar visualization rank distributions between age groups were \taskCl and \taskRe, with all visualizations having percentage point (pp) differences below 25pp. Notable differences in distribution occurred for \taskAn, \taskCo, \taskDi, \taskEx, and \taskRa. From this, we observed that the greatest disagreement in the age groups most typically occurred for the best-performing and worst-performing visualizations, most often revolving around \visLine and \visTable and indicating that those visualizations may be more useful to one group than the other. 

\begin{figure*}[ht]
    \centering
    \captionsetup{type=table}
    \captionof{table}{Percent differences between YA and PLA in median (a) \metAcc and (b) \metTime, per task and best-performing visualization. A larger percent difference indicates greater performance differences between the two groups for that task when using their best-performing visualization.}
    \label{tbl:pct-diff-tv}
        \begin{subfigure}{0.49\textwidth}
        \centering
        \caption{Accuracy}
        \resizebox{0.95\textwidth}{!}{%
        \begin{tabular}{rrrr}
        \hline
        \textbf{Task} & \textbf{YA Vis.} & \textbf{PLA Vis.} & \textbf{\% Diff.} \\ \hline
        \taskAn & \visScatter & \visTable & 1.5\% \\
        \taskCl & \visPie & \visPie & 1.0\% \\
        \taskCo & \visLine & \visScatter & 14.5\% \\
        \taskDe & \visTable & \visTable & 3.4\% \\
        \taskDi & \visBar & \visTable & 10.0\% \\
        \taskEx & \visScatter & \visScatter & 1.4\% \\
        \taskFi & \visTable & \visTable & 3.9\% \\
        \taskOr & \visBar & \visPie & 3.3\% \\
        \taskRa & \visPie & \visTable & 5.3\% \\
        \taskRe & \visTable & \visTable & 0.8\% \\ \hline
        \end{tabular}%
        }
        \label{tbl:pct-diff-tv-acc}
    \end{subfigure}
    \begin{subfigure}{0.49\textwidth}
        \centering
        \caption{Time}
        \resizebox{0.95\textwidth}{!}{%
        \begin{tabular}{rrrr}
        \hline
        \textbf{Task} & \textbf{YA Vis.} & \textbf{PLA Vis.} & \textbf{\% Diff.} \\ \hline
        \taskAn & \visScatter & \visBar & 82.2\% \\
        \taskCl & \visPie & \visPie & 45.6\% \\
        \taskCo & \visLine & \visLine & 86.0\% \\
        \taskDe & \visTable & \visTable & 46.2\% \\
        \taskDi & \visScatter & \visScatter & 24.3\% \\
        \taskEx & \visBar & \visBar & 44.3\% \\
        \taskFi & \visBar & \visBar & 74.9\% \\
        \taskOr & \visBar & \visBar & 43.9\% \\
        \taskRa & \visScatter & \visPie & 33.2\% \\
        \taskRe & \visTable & \visTable & 45.6\% \\ \hline
        \end{tabular}
        }
        \label{tbl:pct-diff-tv-time}
    \end{subfigure}
\end{figure*}

\subsubsection{Time}\text{ }\vspace{0.5em}

\begin{figure*}[ht]
    \centering
    \begin{subfigure}[t]{0.47\textwidth}
        \centering
        \includegraphics[width=\textwidth]{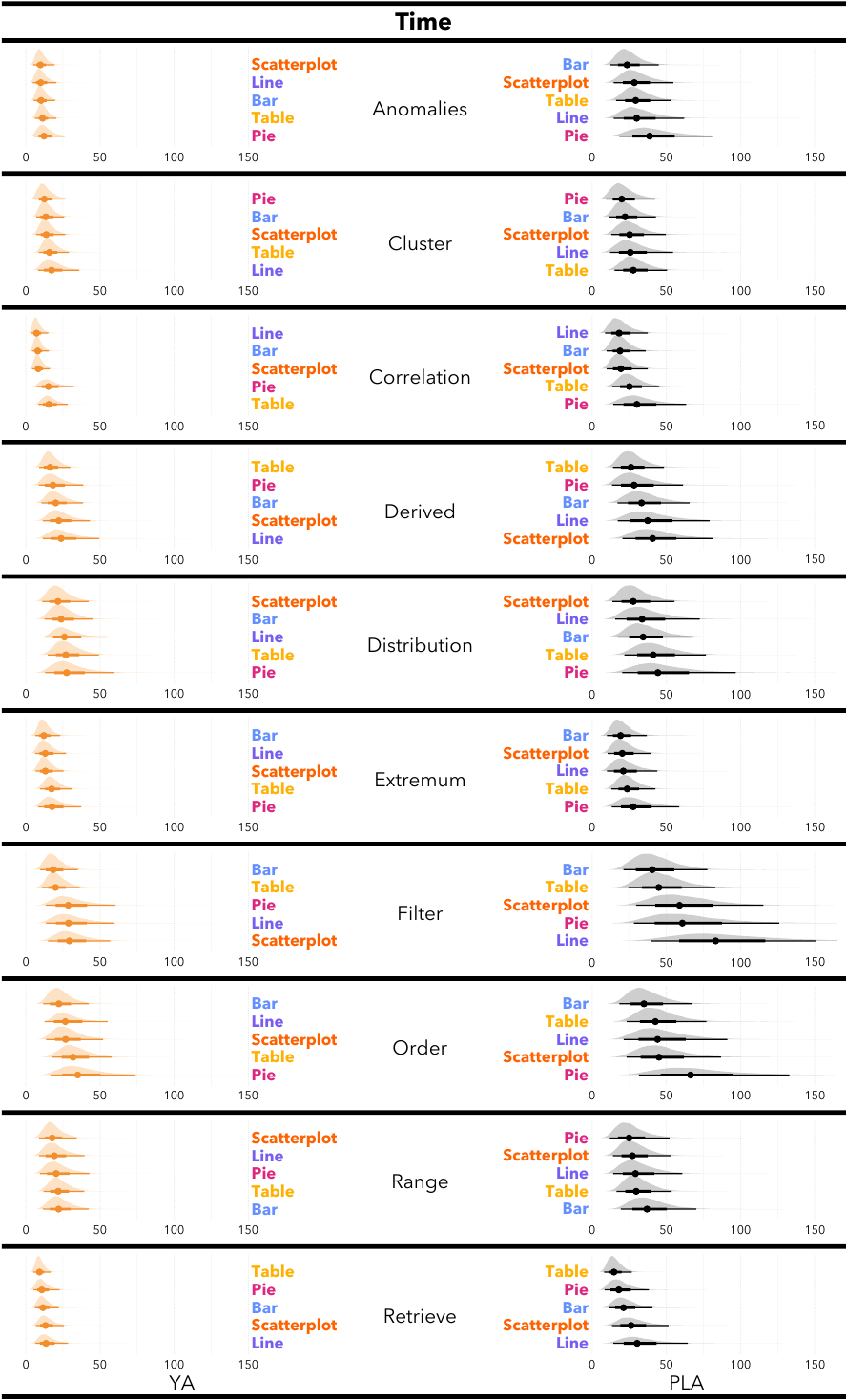}
        \caption{Distribution of mean \metTime for counterfactual YA~\yacbox (left) and PLA~\placbox (right), aggregated by task and visualization. Visualizations for each task are sorted by median from best to worst, and each distribution shows the 50\textsuperscript{th}, 66\textsuperscript{th}, and 95\textsuperscript{th} percentiles.}\label{fig:task-chart-time}
        \Description{Time plots for YA and PLA, per visualization and task. PLA require noticeably more time and have winder ranges.}
    \end{subfigure}
    \hfill
    \begin{subfigure}[t]{0.4935\textwidth}
        \centering
        \includegraphics[width=\textwidth]{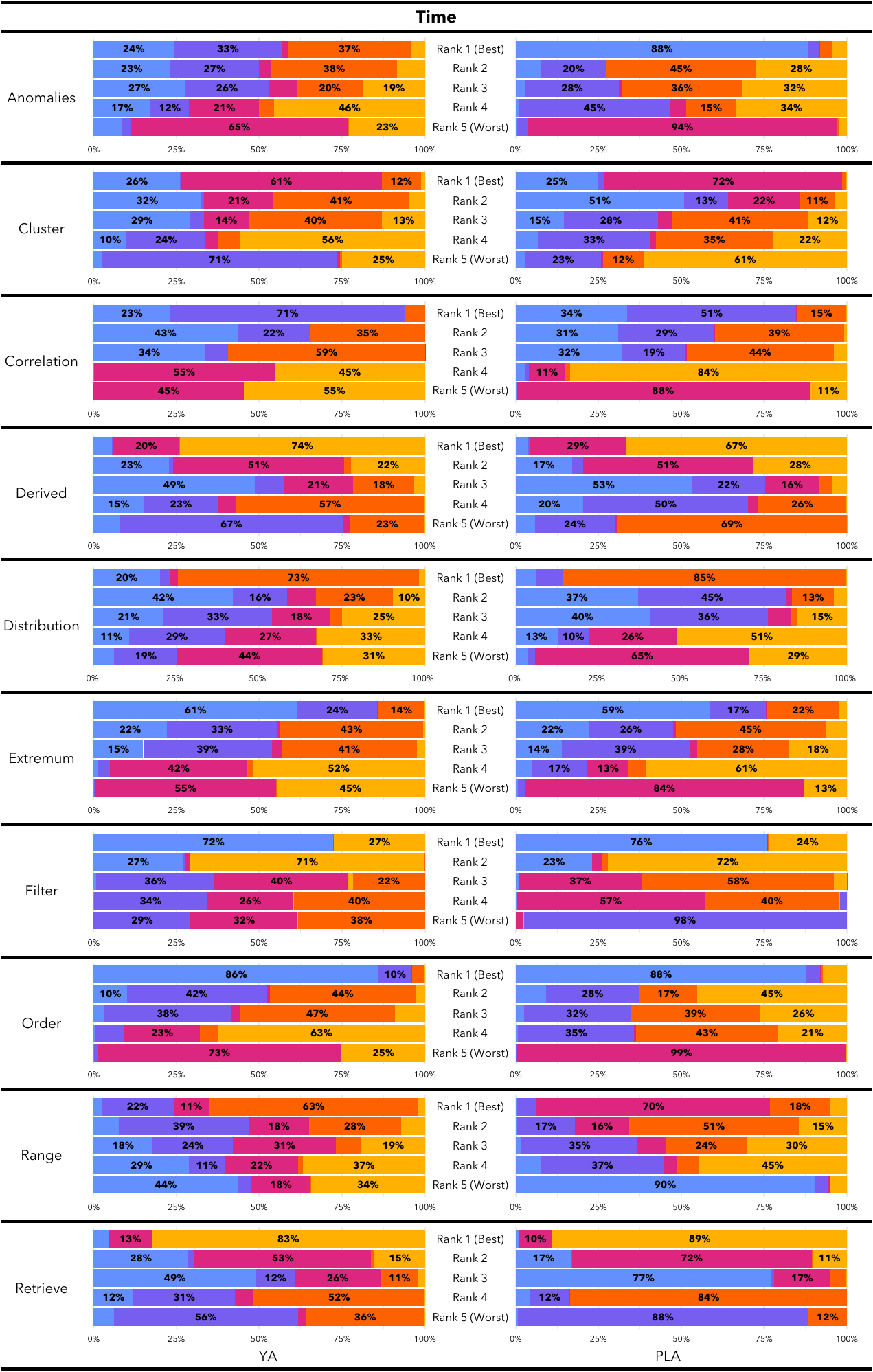}
        \caption{This depicts the proportions of counterfactual participants with each visualization (\visBar~\barcbox, \visLine~\linecbox, \visPie~\piecbox, \visScatter~\scattercbox, and \visTable~\tablecbox) at each rank for time for a given task. YA are depicted on the left, PLA on the right. 
        Any bars without a text label are less than 10\%. 
        }\label{fig:rank-dist-time}
        \Description{Proportions of simulated participants who had each visualization ranked 1-5 for each task for time.}
    \end{subfigure}
    \caption{Task-visualization time results at the (a) aggregate and (b) individual level.}
\end{figure*} 

\textbf{Aggregate Level.}
\autoref{fig:task-chart-time} shows the performance time for each age group at the task and visualization level. For YA, \visBar led to the fastest times for \taskEx, \taskFi, and \taskOr; \visLine for \taskCo, \visPie for \taskCl; \visScatter for \taskAn, \taskDi, and \taskRa; and \visTable for \taskDe and \taskRe. For PLA, \visBar was fastest for \taskAn, \taskEx, \taskFi, and \taskOr; while \visLine was fastest for \taskCo; \visPie for \taskCl and \taskRa; \visScatter for \taskDi; and \visTable for \taskDe and \taskRe. The largest gaps between visualizations for YA occurred for \taskCo with \visPie and \visTable, while for PLA they occurred for \taskFi with \visLine as well as for \taskOr with \visPie.

There was a fair amount of agreement between the two age groups for the ``best'' visualizations for task speed, but there was more disagreement on the ``worst'' visualizations compared to \metAcc.  For the best visualizations, YA and PLA agreed on \taskCl, \taskCo, \taskDe, \taskDi, \taskEx, \taskFi, \taskOr, and \taskRe. However, they disagreed for \taskAn and \taskRa. For the worst-performing visualizations, two groups agreed regarding \taskAn, \taskDi, \taskEx, \taskOr, \taskRa, and \taskRe; they disagreed on \taskCl, \taskCo, \taskDe, and \taskFi. Percent differences of speed (using the best-performing visualization) were much larger than for \metAcc, with the highest differences occurring for \taskAn, \taskCo, and \taskFi. Among all the task-visualization pairs, the largest percent differences between the age groups occurred for \{\taskAn, \visPie\}, \{\taskAn, \visLine\}, \{\taskFi, \visLine\}, and \{\taskAn, \visScatter\}. While \metAcc had major rank disagreements for certain tasks, \metTime had no disagreement beyond 2 ranks. This indicates a stronger general agreement between the two groups regarding the relative aptitude of visualizations for completing tasks quickly.

\textbf{Individual Level.} 
As with \metAcc, we then counted the proportions of counterfactual participants that performed fastest (Rank 1) to slowest (Rank 5) with each visualization for each task. Notably, both age groups had at least one rank with a majority for each task, as opposed to \metAcc. YA had a majority agreement for the ``best'' visualization for all tasks but \taskAn, while PLA did for all tasks. On the other hand, YA had a majority ``worst'' visualization for all tasks but \taskDi, \taskFi, and \taskRa, while PLA did for all tasks. Similar to \metAcc, the lack of majority from YA for some tasks compared to PLA having all majorities for both ``best'' and ``worst'' may indicate slightly more heterogeneity in the YA population at the task-visualization level.

The tasks \taskEx, \taskDi, and \taskRe comparatively have the most-similar visualization rank distributions between the two age groups, all with percentage point differences roughly around or below 25pp; however, these differences are notably larger than the equivalently most-similar tasks for \metAcc. The largest differences in proportions occurred for \taskAn, \taskCl, \taskCo, \taskDe, \taskFi, \taskOr, and \taskRa. Consistent with \metAcc, most large disagreements in rank distribution between the groups were at the extremes (ranks 1 and 5), however unlike \metAcc they did not just primarily involve \visLine and \visTable. 

\section{Discussion}
Recent work at the intersection of aging and HCI encourages moving away from the \textit{discourse of deficits}, which focuses on design for PLA that considers the aging-related changes in perception, cognition, and physical abilities~\cite{vines2015age}. However, that creates a nebulous situation for those working in GerontoVis, as the broader visualization field often uses perceptual measures (\iDotE accuracy, speed) and psychophysical studies to both evaluate the effectiveness of a visualization~\cite{burns2020evaluate} and create design recommendations~(\eDotG~\cite{shao2024does}), though other evaluation measures are becoming increasingly common (\eDotG~\cite{lan2023affective, borkin2015beyond, rogha2024impact}). Hence, we aim to discuss our results using a middle-ground approach that considers the effects of physiological changes on performance while avoiding language that create and perpetuate negative stereotypes about PLA and the aging process. As noted by While et al.~\cite{while2024gerontovis}, much of our empirical understanding of visualization effectiveness is through the lens of a younger population of study participants, so while we may describe how PLA \textit{differ} from YA, it is important to clarify that YA are not necessarily the ``standard'' that PLA should conform to through design choices. Instead, these results should allow us to create visualizations that are usable by a broader audience with more varying needs.

\subsection{How Does Aging Interact With Analysis Task and Visualization Type?}
The performance variations revealed through our analysis prompt a nuanced consideration of the similarities and differences in experienced task difficulty between YA and PLA. In this section, we further discuss our findings' implications and speculate how age, visualization, and task can impact visualization performance.

\textbf{Aging More Noticeably Impacts Specific Tasks and Visualizations.} 
At the task level, PLA encountered greater relative difficulty with \taskFi compared to YA. This may be due to declines in visual working memory~\cite{faubert2002visual, klencklen2017working, faubert2002aging} (VWM) and visual acuity~\cite{mitzner2015considering} that can occur with age, as that task requires searching, computing, and remembering information. 
Meanwhile, PLA found \taskAn noticeably more time-consuming than other tasks in comparison to YA. Because this task required assessing which data point had an unexpected value, PLA may have taken more time to study the options and reduce the likelihood of errors~\cite{endrass2012speeding}. Furthermore, finding an anomalous data point among all possible options may have taken longer due to changes in VWM and visual acuity.
Among the five visualizations, a majority of PLA achieved their lowest accuracy with \visLine for several tasks and rarely had it as their top-ranked task. We speculate that this reduced accuracy results from age-related changes in visual acuity, leading to difficulty finding and following the line when identifying data values.  Meanwhile, PLA achieved their highest accuracy with \visTable noticeably more often than YA. This preference for numerical representations is consistent with previous work by While et al.~\cite{while2024glanceable} where PLA voiced a preference for numerical displays on smartwatches instead of visualizations. Additionally, work in psychology has observed numerical and arithmetic computation speed from PLA that, depending on the task, can be slower or faster than YA~\cite{gearysimple}. Further work is required to more deeply understand why some tasks and visualizations are more noticeably impacted with increasing age.

\textbf{More Examination is Needed for Time-Limited Scenarios.}
Our findings indicate that with aging, the pace of visual data analysis may reduce, though accuracy remains largely unaffected. This aligns with prior perceptual psychology studies indicating that visual acuity remains relatively stable with age in healthy individuals~\cite{endrass2012speeding, faubert2002aging} while the speed of visual information processing often declines~\cite{guest2015aging}. Similarly, Le et al.~\cite{le2014elementary} and While et al.~\cite{while2024glanceable} found that PLA took more time than YA to complete visual analysis tasks but achieved comparable accuracy. We concur with Le et al.~\cite{le2014elementary} in that the similarities in accuracy may be due in large part to the additional time spent, perhaps compensating for age-related declines in the perception of geometric properties such as orientation, symmetry, parallelism, motion direction, and collinearity that are critical for interpreting visualizations~\cite{meng2019age, ball1986improving, faubert2002visual}.
PLA's lower pace may also reflect compensatory strategies to avoid errors, as suggested by Endrass et al.~\cite{endrass2012speeding}. However, such strategies may be impractical in time-critical contexts like clinical decision-making, where time pressure can impair visual performance and decision-making regardless of age~\cite{rieger2022understanding,rieger2022human, rieger2021visual}. Double encodings, reference structures, and annotations may help PLA process visualizations faster, however the necessity of amplifying analysis speed highly depends on the context~\cite{while2024glanceable}. More work is needed to make sure that, given limited time, PLA are still able to make accurate data-driven decisions.

\textbf{Aging Has a Multifaceted Impact on Performance.}
In the task-based analysis, we observed that PLA showed greater heterogeneity in the relative ranking of tasks for both \metAcc and \metTime, whereas YA showed strong agreement toward single (disjoint) sets for their three best-, four middle-, and three worst-performing tasks. This comparative heterogeneity across both tasks aligns with observations in broader gerontology~\cite{jaul2021characterizing}, resulting from the wide variety of ways that changes in aging can manifest over time and impact task performance~\cite{nguyen2021health}. 
Similarly, the distributions of mean \metTime values for PLA at both the task level and the task-visualization level were noticeably wider than YA, however this was not the case for \metAcc. Because \metAcc between the age groups at both the task and task-visualization level showed similar values and variance, this suggests that PLA's heterogeneity manifests moreso in their visual analysis speed. This suggests that changes due to aging may, overall, be more impactful on aspects affecting speed.
However, we interestingly observed that the distribution of visualizations leading to the best performance was more evenly spread among YA than PLA. PLA, in contrast, exhibited greater consensus, with majority or near-majority agreement on a specific visualization at each rank for many tasks across \metAcc and \metTime. We speculate that this stronger agreement among PLA regarding the ``best'' visualization for many tasks may stem from the reduced feasibility of certain visualizations' design aspects due to age-related changes. This likely results in a greater ``convergence'' toward a narrower set of viable visualizations for each task. This set of results leaves us with an interestingly contradictory scenario that requires further study: PLA appear to exhibit greater heterogeneity in task-level performance as a result of changes due to aging, however these changes may actually lead to greater agreement on the relative usability of certain visualizations over others for some tasks.

\subsection{What Are The Resulting Design Implications for PLA?}
Leveraging the results from our Bayesian analysis, we now detail the resulting set of design recommendations for designers aiming to better-support PLA as consumers of visual data. 

\textbf{Offer Additional Numerical Representations.}
Due to PLA's pronounced higher performance with \visTable, we recommend incorporating more numerical representations of data, either as a redundant encoding~\cite{kong2012graphical} or via a tabular representation of the data. In fact, the latter falls in line with existing recommendations for as data tables as an accessibility measure~\cite{wang2022makes}. For example, a Bar chart could have numerical labels at the top of the bars indicating their value, facilitating faster reading and analysis. It is also important to consider the low-level tasks that PLA may be using, \eDotG \visTable was not best-suited for the \taskCo task; furthermore, if screen space is limited or the visualization is already visually busy, consider making it customization option to reduce the possibility of creating screen clutter.  Text added in this way should conform to WCAG AAA standards, which includes a 4.5:1 contrast ratio with the background and a font size of at least 18 pixels~\cite{wcag2018}.

\textbf{Supplement Line Charts for Greater Usability.}
Our analysis illustrated that \visLine was rarely well-suited for supporting PLA in terms of accuracy, however it was top-ranked for \taskCo for a majority of PLA in terms of time and was also one of the most common top-ranked visualizations for \taskDi and \taskEx in terms of time. Thus, this visualization can be beneficial, however it needs to be adjusted slightly to increase accuracy for PLA. One simple recommendation would be to allow the user to customize the width of the line to their liking, which would reduce the impact of lowered visual acuity. If space and visual clutter amount allows it, another recommendation would be to augment Line charts with visual aids such as gridlines~\cite{bartram2011gridlines} or Scatterplot-style marks at each point; for more dense datasets, the marks could be added at evenly-spaced intervals or points of interest. During analysis, we observed that \visScatter was one of the top-ranked for several tasks, illustrating its possible feasibility as a redundant encoding. Lastly, incorporating the above-mentioned numerical representations as a redundant encoding with the Scatterplot marks could also improve visual analysis performance. If the marks will be interactive, they should at least confirm to WCAG AA requirements of at least $24 \times 24$ pixels, however if space allows it then we recommend the WCAG AAA size of $44\times 44$ pixels. 

\textbf{Provide Annotations to Reduce Memory Overhead.}
In the previous section we partly attributed lower performance for two tasks (\taskFi and \taskAn) to possibly-reduced visual working memory (WVM). To reduce its possible impact, we recommend providing built-in annotations to visualizations~\cite{borkin2015beyond}, \eDotG pointing out extrema or anomalous data if they are known beforehand; if expected user tasks are unknown at the time of design, then providing a simple user interface to allow users to add their own annotations could allow PLA to provide their input. This would be especially helpful for temporal data tracking an important metric such as blood pressure, heart rate, or sleep quality. Consistent with the above recommendations, the annotations may benefit from including numerical representations depending on the situation, and annotations when possible should also abide by the WCAG AAA recommended contrast ratio and font size for maximum usability.

\textbf{Provide Options for Personalized Visualizations.}
We note that this design implications section will not be providing task-based rankings of visualizations in the same style as the prior study, as recent work has cautioned against the use of strict rankings as to guide design decisions due to the influence of individual differences~\cite{davis2022risks}. The increasing evidence against a \textit{one size fits all} solution to visualization design, combined with the known heterogeneity of the PLA population, motivates our recommendation to instead provide multiple visualization options that PLA can select from. If space or implementation complexity limit the number visualizations possible, then our recommendation would be to aim for population coverage, \iDotE choosing the visualizations that led the largest percentages of PLA to achieving their best performance (further detailed in \autoref{subsec:task-vis-performance}). This recommendation aligns with recent work advocating for personal visualization~\cite{huang2014personal,bullock2022let} that can allow for customization that promotes improved visual analysis.

\section{Limitations}
As a conceptual replication, this study inherits the limitations of the prior study~\cite{saket2018task}. This work only covered a set of 5 basic visualizations used in conjunction with 10 low-level visual analysis tasks, whereas differences as a result of aging may vary greater when the visualizations and tasks become more complicated due to the amount and complexity of presented information. For example, it is unclear if \visTable, which PLA noticeably performed better with for multiple tasks, would still be as applicable in those scenarios.  Additionally, our study compared the accuracy and time performance of YA and PLA, which can demonstrate some perceptual capabilities but are not sufficient for measuring other insightful metrics such as how much participants learned from the visualizations~\cite{burns2020evaluate}. We measured accuracy as a binary using multiple choice questions with four answers, which leaves out some granularity of understanding in comparison to other studies that ask questions requiring numerical estimation and more fine-grained responses~(\eDotG~\cite{davis2022risks}). 

Furthermore, the study was conducted with online-recruited users viewing visualizations on a computer screen, whereas PLA may encounter visualizations on a variety of other devices with widely-varying screen sizes~\cite{while2024gerontovis}.  It is unclear whether these results would translate to a smaller device such as a tablet, or even smaller devices with different screen aspect ratios such as a cellphone or smartwatch. As an online study that required participants to be located in the US in order to participate, it is uncertain whether these results would generalize to PLA in other countries, and further work is needed to explore the diverse visualization experiences and design considerations of PLA in a greater variety of locations. We did not ask participants whether they used accessibility tools such as screen readers or screen magnifiers during the study, which may have had an impact on their performance; additional research is needed to understand how visualization performance interacts with these tools for PLA, as well as how effective design can work in tandem with them to create a more fluid visual analysis experience for PLA.
The study did not evaluate the impact of specific design choices such as contrast polarity, text annotations, and line width, which additionally would allow for greater depth of understanding.

Lastly, there are intrinsic challenges when conducting a Bayesian analysis. For each performance metric (\metAcc and \metTime), we applied the same prior across all combinations of tasks and visualizations, which may not accurately depict their underlying differences in performance. While the weakly-informed priors allowed the large amount of collected data to have the greatest impact on the likelihood, giving more flexibility to the model through more nuanced priors may lead to somewhat different results. These models' complexity can encode a multitude of underlying patterns in data, so further analysis with these models could generate an even greater understanding of how aging interacts with tasks and visualization type; collecting more data while using a more informed prior~\cite{kay2016researcher} or comparing correlations in \metAcc and \metTime between various visualizations and tasks~\cite{davis2022risks}, for example, could lead to further insights into the relationship between aging and visualization performance. Using the models to generate counterfactual participants as opposed to only using the raw participant data certainly has the capability for introducing bias and noise into the analysis, however we did our best to reduce its impact on our takeaways. This included comparing the model's results to the empirical data to check that they roughly match (supplementary materials), filtering the data for outliers to prevent unintended model biasing (\autoref{sec:data-prep}), and testing each prior's robustness in a sensitivity analysis (\autoref{sec:model-desc}).

\section{Conclusion and Future Work}

This work presents the results of a crowdsourced study and Bayesian analysis investigating the performance and design of visualizations for people in late adulthood (PLA). Our analysis found that aging affects visual analysis speed for PLA while achieving accuracy similar to younger adults (YA), sometimes using different visualizations.  

Future work continuing this research can investigate the impact of aging on the use of more complex visualizations and analysis tasks, \eDotG comparing and contrasting YA's and PLA's performance using small multiples, graphs, and trees. It is also important to assess the impact of other visualization design choices such as graph stroke width and shape, graph orientation, and existing accessibility technologies such as screen readers on the performance of PLA in a visualization context. These can open the door to new directions of study within GerontoVis covering the impact of various design choices that can enrich our understanding of how aging affects visualization performance, allowing for more age-inclusive visualization design. Another avenue would be investigating the impact of age-related decline in fine motor skills on the ability of PLA to use interactive visualizations. It would also be beneficial to investigate other metrics of ``successful'' visualization use and ``good'' visualization design for PLA beyond accuracy and speed. Further discussion is needed in the visualization community as to how we ultimately want to frame our approach to visualization research with PLA, keeping in mind the broader HCI community's warnings about deficit-driven design. These paths can lead us to understanding more of humanity's experiences with visualization and bringing more people to the table for visual data analysis.

\begin{acks}
The authors wish to thank Dr. G. Jay Kerns, Professor of Mathematics and Statistics at Youngstown State University, for his valuable guidance and feedback during the course of developing the Bayesian models. We would also like to thank Dr. Bahador Saket for providing detailed access to prior study's software, datasets, and collected data. Lastly, we want to thank Mohammad Hadi Nezhad for his assistance in running our version of the online study.
\end{acks}

% \newpage
\bibliographystyle{ACM-Reference-Format}
\bibliography{source/manuscript}

\end{document}